\documentclass[12pt,preprint]{aastex}

\usepackage{wasysym}

\shorttitle{}
\shortauthors{Nesvorn\'y et al.}

\begin{document}
\baselineskip 19.pt

\title{The Role of Early Giant Planet Instability in \\the Terrestrial Planet Formation}

\author{David Nesvorn\'y$^1$, Fernando V. Roig$^2$, Rogerio Deienno$^1$}
\affil{(1) Department of Space Studies, Southwest Research Institute,\\
1050 Walnut St., Suite 300, Boulder, CO 80302, United States}
\affil{(2) Observat\'orio Nacional, Rua Gal. Jose Cristino 77,\\ Rio de Janeiro, RJ 20921-400, Brazil}

\begin{abstract} 
The terrestrial planets are believed to have formed by violent collisions of tens of lunar- to Mars-size protoplanets 
at time $t<200$ Myr after the protoplanetary gas disk dispersal ($t_0$). The solar system giant planets rapidly formed 
during the protoplanetary disk stage and, after $t_0$, radially migrated by interacting with outer disk planetesimals. 
An early ($t<100$ Myr) dynamical instability is thought to have occurred with Jupiter having 
gravitational encounters with a planetary-size body, jumping inward by $\sim0.2$-0.5 au, and landing on its current, 
mildly eccentric orbit. Here we investigate how the giant planet instability affected formation of the terrestrial 
planets. We study several instability cases that were previously shown to match many solar system constraints. We find 
that resonances with the giant planets help to remove solids available for accretion near $\sim1.5$ au, thus stalling 
the growth of Mars. It does not matter, however, whether the giant planets are placed on their current orbits at $t_0$ 
or whether they realistically evolve in one of our instability models; the results are practically the same. The tight 
orbital spacing of Venus and Earth is difficult to reproduce in our simulations, including cases where bodies grow from 
a narrow annulus at 0.7-1 au, because protoplanets tend to radially spread during accretion. The best results are obtained 
in the narrow-annulus model when protoplanets emerging from the dispersing gas nebula are assumed to have (at least) the 
Mars mass. This suggests efficient accretion of the terrestrial protoplanets during the first $\sim10$ Myr of the 
solar system.
\end{abstract}

\section{Introduction}

In the standard model of the terrestrial planet formation (Wetherill 1990), accretional collisions between 1 to 1000 km 
planetesimals lead to gradual build up of lunar- to Mars-size protoplanets, which gravitationally interact and further grow during 
a late stage of giant impacts (e.g., Chambers \& Wetherill 1998). Indeed, radiometric data indicate that the Moon-forming impact
on proto-Earth happened relatively late, some $\sim$30-150 Myr after the appearance of the first solar system solids (see Kleine 
\& Walker 2017 and Canup et al. 2020 for reviews). $N$-body simulations of the late-stage accretion are required to match various 
constraints (see Morbidelli et al. 2012 for a review), including the similar semimajor axes of Venus and Earth (only 
$\simeq0.28$ au radial separation), the small masses of Mercury and Mars, and the angular momentum deficit (AMD) of planets -- 
a measure of their orbital excitation. The results indicate that the terrestrial planets could have accreted from a narrow annulus 
(0.7-1 au; Hansen 2009), perhaps because the protoplanet/planetesimal disk was truncated at $\sim 1$ au by Jupiter's inward and then 
outward migration in a gas disk (Walsh et al. 2011). Alternatively, the early growth of terrestrial protoplanets has been regulated 
by accretion of small, pebble-size particles (Johansen et al. 2015), rather than larger planetesimals, and small Mars 
reflects the inefficiency of pebble accretion beyond $\sim 1$ au (Levison et al. 2015, Izidoro et al. 2015).       

Clement et al. (2018) suggested that Mars's growth was stunted by the early migration/instability of the giant planets, which 
presumably excited and depleted the population of protoplanets at $\sim1.5$ au. The resonant structure of the Kuiper 
belt provides clear evidence for the planetesimal-driven migration of the giant planets (Hahn \& Malhotra 2005). A smooth migration, 
however, would be incompatible with the dynamical structure of the asteroid belt and mildly eccentric orbits of the giant planets: 
thus the need for a dynamical instability (Tsiganis et al. 2005, Morbidelli et al. 2010). The instability was originally thought 
to have happened hundreds of Myr after the solar system birth (Gomes et al. 2005, Bottke et al. 2012), at which point the terrestrial 
planet system was already in place. This turned out to be a problem because the instability was shown to excessively excite 
the terrestrial planet orbits (Agnor \& Lin 2012, Kaib \& Chambers 2016; also see Brasser et al. 2013 and Roig et al. 2016). 
Recent evidence suggests that the instability happened early, within the first $\sim$100 Myr, and possibly only a few Myrs after 
the gas disk dispersal (Nesvorn\'y et al. 2018, Morbidelli et al. 2018, de Sousa et al. 2020). This would imply that the giant planet 
migration/instability happened {\it during} the terrestrial planet formation (Clement et al. 2018; also see Walsh \& Morbidelli 2011). 

Here we perform $N$-body simulations to establish the effect of early giant-planet instability on the terrestrial planet 
formation. We use several instability cases that were previously shown to match many solar system constraints (see Nesvorn\'y 
2018 for a review), and different initial distributions of planetesimals/protoplanets at 0.3-4 au. The goal is to show how these 
`standard' instability models affect the terrestrial planet formation and how the results compare to those obtained from a much 
broader set of instability cases investigated by Clement et al. (2018). We find that the details of giant planet instability do 
not matter with small Mars potentially forming even in the case where the giant planets are placed on their present orbits at
the beginning of simulations. The simulations with a radially extended disk of protoplanets, however, fail to match the tight 
orbital spacing of Venus and Earth (also see Deienno et al. 2019). To obtain the correct spacing, the terrestrial protoplanets 
need to start in a narrow annulus (0.7-1 au; Hansen 2009) and have (at least) the Mars mass to begin with (Jacobson \& Morbidelli 2014). 
The method and initial conditions are described in Sect. 2, the results are reported in Sect. 3. and summarized in Sect 4.
 
\section{Method}

\subsection{Initial distribution of protoplanets and planetesimals}

Our $N$-body simulations start at the time of the gas disk dispersal, $t_0$. In the standard setup, there are $N_{\tt pp}=100$ 
protoplanets and $N_{\rm pm}=1000$ planetesimals, each representing the total mass of 2 $M_{\rm Earth}$ (i.e., the total mass of 
solids is $M_{\rm solid}=4$ $M_{\rm Earth}$; each protoplanet has the mass $M_{\rm pp} \simeq 1.6$ $M_{\rm Moon}$ and each planetesimal 
has the mass $M_{\rm pm} \simeq 13$ $M_{\rm Ceres}$, where $M_{\rm Earth}=5.97 \times 10^{27}$ g, 
$M_{\rm Moon}=7.35 \times 10^{25}$ g and $M_{\rm Ceres}=9.38 \times 10^{23}$ g). 
The bulk density of protoplanets and planetesimals is set to $\rho=3$ g cm$^{-3}$. The initial eccentricities and initial inclinations 
are chosen to follow the Rayleigh distributions with $\sigma_e=0.005$ and $\sigma_i=0.0025$. The protoplanets and planetesimals 
are distributed with the surface density profile $\Sigma(r)=r^{-\gamma}$, where $r$ is heliocentric distance, in a disk between 
$r_{\rm in}$ and $r_{\rm out}$, with $\gamma=1$ in the standard case.

We test three different disk setups. In the first case, both the protoplanet and planetesimal disks have $r_{\rm in}=0.3$ au 
and $r_{\rm out}=4$ au. This means that $\sim1$ $M_{\rm Earth}$ in protoplanets\footnote{Here we prefer to use the term 
`protoplanets' rather than `planetary embryos' but both terms just mean `large bodies' from which the terrestrial 
planets are assembled. The protoplanets are assumed to grow by planetesimal and pebble accretion before $t_0$ (e.g., Wetherill 1990, 
Johansen et al. 2015).} is placed into the asteroid belt region. In the second case, the protoplanets are distributed in 
a narrow annulus between $r_{\rm in}=0.7$ au and $r_{\rm out}=1$ au, whereas the planetesimal disk follows the same distribution as 
in the first setup (i.e., $r_{\rm in}=0.3$ au and $r_{\rm out}=4$ au). This would approximate the state of the inner solar system 
at $t_0$ in the Grand Tack model (Walsh et al. 2011), except that the planetesimals in the asteroid belt region start on low 
eccentricity/inclination orbits (see Sect. 3.4 for simulations with excited asteroid orbits). In the 
third case, protoplanets and planetesimals occupy separate radial intervals: $r_{\rm in}=0.3$ au and $r_{\rm out}=1.5$ au for 
protoplanets, and $r_{\rm in}=1.5$ au and $r_{\rm out}=4$ au for planetesimals. This case is motivated by the results of Levison et 
al. (2015), who showed that the protoplanet and planetesimals disks may not overlap at $t_0$ (also see Walsh \& 
Levison 2019 and Clement et al. 2020b).\footnote{Our setup is arguably a very simplified representation of the initial conditions expected from 
Levison et al. (2015). For example, the protoplanets in Levison et al. (2015) have a steep radial distribution with 
$\gamma \sim 3$-5 (also see Izidoro et al. 2015), whereas here we use $\gamma \sim 1$ and truncate the protoplanet disk at 
1.5 au.} The three cases discussed above are referred to as C98 (after Chambers \& Wetherill 1998), W11 (after Walsh et al. 2011) 
and L15 (after Levison et al. 2015).

We consider six different dynamical models for each protoplanet/planetesimal distribution. In the first model, there are 
no giant planets whatsoever. This is clearly unrealistic, because the giant planets formed before the terrestrial planets, but
it is useful to consider this case as a reference with no external influence on the terrestrial planet formation. In the second 
model, we start the simulations with the four giant planets (Jupiter to Neptune) on their current orbits. Note that this would be 
equivalent to assuming that the giant planets evolved to their current orbits before $t_0$ and there was no planetesimal-driven migration and 
dynamical instability after $t_0$. Again, this `static' case is useful to consider as a benchmark for models
with the planetary migration/instability (Sect 2.2). In the third model, we extract the initial configuration of giant planets from 
the instability Case 1 reported in Nesvorn\'y et al. (2013). In Case 1, five giant planets (Jupiter, Saturn and three ice giants) 
start on orbits in a compact resonant chain with the period ratios following the sequence 3:2,3:2,2:1,3:2 (i.e., Jupiter and Saturn are 
in the 3:2 mean motion resonance). Since we do not consider the outer planetesimal disk, however, the giant planets do not migrate in this 
model. This would corresponds to a case where the terrestrial planet formation was completed prior to the migration/instability 
of the giant planets. The effects of the giant planet migration/instability on a fully formed terrestrial planet system were 
studied in Brasser et al. (2009, 2013), Agnor \& Lin (2012), Kaib \& Chambers (2016) and Roig et al. (2016). The three models described above 
are denoted {\tt nojov}, {\tt jovend} and {\tt jovini}, where `jov' stands for the jovian planets. The three remaining models are described 
below.

\subsection{Giant planet migration/instability}

We take advantage of our previous simulations of the giant planet migration/instability (e.g., Nesvorn\'y \& Morbidelli 2012, 
hereafter NM12; Deienno et al. 2017) and select three cases. The three models started with fully-formed giant planets at $t_0$. 
The giant planets were assumed to have orbits in mutual resonances that have been presumably established during the previous stage 
of convergent migration in a gas disk (Masset \& Snellgrove 2001, Pierens \& Nelson 2008, Pierens \& Raymond 2011). In all three 
cases, five giant planets began in the resonant chain (3:2,3:2,2:1,3:2), a configuration that was previously shown to work best 
to generate plausible dynamical histories (NM12; Deienno et al. 2017). The additional ice giant had mass comparable to that of Uranus/Neptune. 
A massive outer disk (mass $M_{\rm disk}$) of planetesimals was placed beyond the outermost ice giant. The dynamical evolution of the 
giant planets and planetesimals was then tracked using the $N$-body integrator known as {\tt SyMBA} (Duncan et al. 1998). 
 
\begin{figure}
\epsscale{0.6}
\plotone{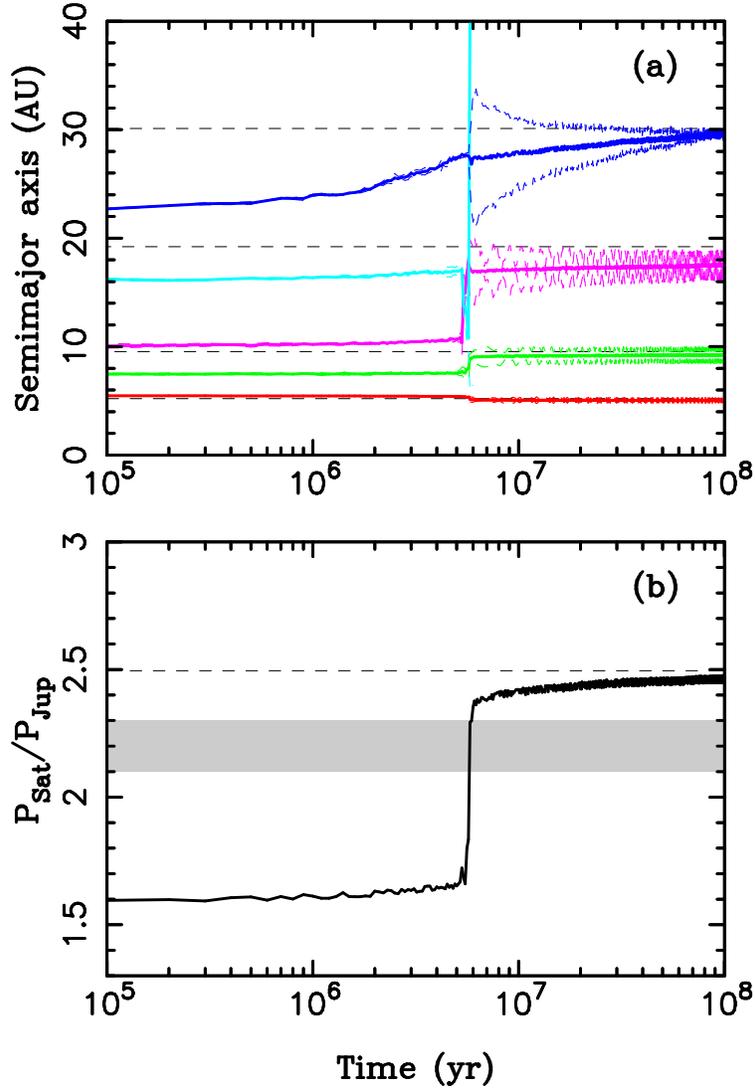}
\caption{The orbital histories of the giant planets in {\tt case1}. The planets started in the (3:2,3:2,2:1,3:2) resonant chain.
The outer disk of planetesimals, not shown here, with the total mass $M_{\rm disk}=20$ M$_{\rm Earth}$ was placed beyond Neptune. 
Panel a: The semimajor axes (solid lines), and perihelion and aphelion 
distances (dashed lines) of each planet's orbit.  The black dashed lines show the semimajor axes of planets in the present Solar System. Panel 
b: The period ratio $P_{\rm Sat}/P_{\rm Jup}$, where $P_{\rm Jup}$ and $P_{\rm Sat}$ denote the orbital periods of Jupiter and Saturn.
The dashed line shows $P_{\rm Sat}/P_{\rm Jup}=2.49$ corresponding to the period ratio in the present solar system. The shaded 
area approximately denotes the interval where the secular resonances with the terrestrial planets occur (Brasser et al. 2009, 
Agnor \& Lin 2012, NM12). The third ice giant is ejected from the solar system during the instability at $t \simeq 5.8$ Myr.}
\label{fig1}
\end{figure}

\begin{figure}
\epsscale{0.6}
\plotone{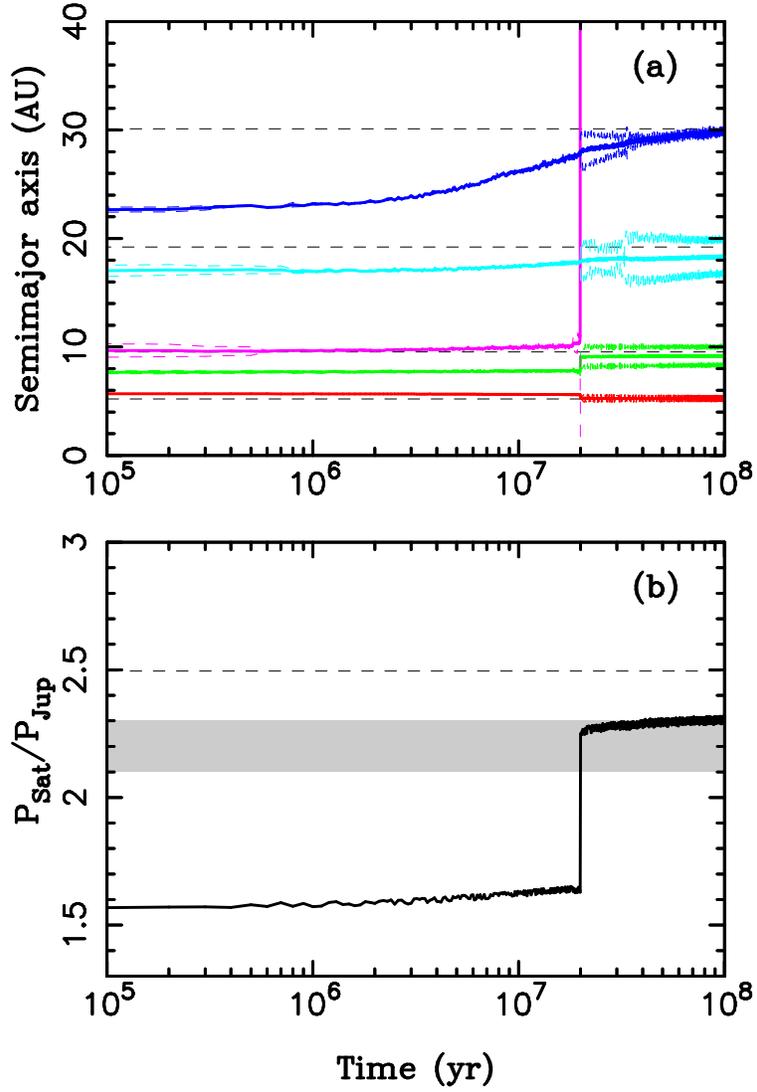}
\caption{The orbital histories of the giant planets in {\tt case2}. See caption of Fig. \ref{fig1} for the figure description.
Here, the  third ice giant is ejected at $t \simeq 20$ Myr.}
\label{fig2}
\end{figure}

The properties of two of our instability models, hereafter {\tt case1} and {\tt case2}, are illustrated in Figs. \ref{fig1} and 
\ref{fig2}. The {\tt case1} model is identical to Case 1 used in Nesvorn\'y et al. (2013, 2014, 2017), Deienno et al. (2014) 
and Roig \& Nesvorn\'y (2015). This model satisfies many solar system constraints including the present orbits of the giant planets 
themselves, the number and orbital distribution of Jupiter Trojans and irregular moons of the giant planets, the dynamical 
structure of the asteroid and Kuiper belts, etc. (see Nesvorn\'y 2018 for a review). Moreover, Roig et al. (2016) demonstrated, 
assuming that the terrestrial planets were already in place when the instability happened, that the {\tt case1} model could 
explain the excited orbit of Mercury and the AMD of the terrestrial planet system. 

Our {\tt case2} is similar to {\tt case1} but the jump of Jupiter (and Saturn) is not as large as in {\tt case1} (Fig. \ref{fig2}). 
In {\tt case2}, the terrestrial planet region is expected to experience a stronger coupling to the giant planets via sweeping 
secular resonances (e.g., Brasser et al. 2009, Agnor \& Lin 2012). Here we follow the general suggestion of Clement et al. (2018) that a broader 
range of instabilities should be considered, but we do not want to distance ourselves too much from the class of instabilities 
that work reasonably well for other solar system constraints (Nesvorn\'y 2018). For the third instability model considered 
here, {\tt case3}, we refer the reader to Figs. 1 and 3 in Deienno et al. (2018), who showed that a more complex dynamical 
evolution of Jupiter and Saturn during the instability could explain the excited orbits of main belt asteroids (even if 
asteroids had $e \sim i \sim 0$ at $t_0$). This contrasts with {\tt case1} where the instability itself does not sufficiently excite 
asteroid orbits (Roig \& Nesvorn\'y 2015, Nesvorn\'y et al. 2017). In {\tt case1}, the orbital excitation of main
belt asteroids would have to predate the onset of giant planet migration/instability. 

The final orbital period ratio of Jupiter and Saturn is $P_{\rm S}/P_{\rm J} \simeq 2.3$-2.5 in the three instability models 
considered here (e.g., Figs. \ref{fig1} and \ref{fig2}; $P_{\rm S}/P_{\rm J}=2.49$ in the real solar system). This can be compared 
to the very broad variation of final $P_{\rm S}/P_{\rm J}$ in the models considered in Clement et al. (2020). The final amplitude of 
Jupiter's eccentric mode is $E_{\rm 55} \simeq  0.025$-0.042 in our three models (highest in {\tt case2}, lowest in {\tt case1}).
These values are somewhat lower than $E_{\rm 55} \simeq  0.044$ of the real Jupiter's orbit.
 
In all three dynamical models considered here, the early solar system is assumed to have five giant planets (Jupiter, Saturn and 
three ice giants). This is because NM12 showed that various constraints, such as the final orbits of outer planets, can most 
easily be satisfied if the solar system started with five giant planets and one ice giant was ejected during the instability 
(Nesvorn\'y 2011, Batygin et al. 2012, Deienno et al. 2017). The case with four initial planets requires a massive planetesimal 
disk to avoid losing a planet, but the massive disk also tends to produce strong dynamical damping and long-range residual migration 
of Jupiter and Saturn that violate constraints. It is therefore difficult to obtain a plausible evolution history starting with 
four planets. 

A shared property of the selected models is that Jupiter and Saturn undergo a series of planetary encounters with the ejected 
ice giant. As a result of these encounters, the semimajor axes of Jupiter and Saturn evolve in discrete steps. While the 
semimajor axis can decrease or increase during one encounter, depending on the encounter geometry, the general trend is such 
that Jupiter moves inward, i.e. to shorter orbital periods (by scattering the ice giant outward), and Saturn moves outward, i.e., to 
longer orbital periods (by scattering the ice giant inward). This process leads to the dynamical evolution known as the jumping-Jupiter 
model,\footnote{The jumping-Jupiter model was originally thought, assuming the late instability model (e.g., Gomes et al. 2005, 
Levison et al. 2011), to be required from the terrestrial planet constraint (Brasser et al. 2009, Walsh \& Morbidelli 2011, 
Agnor \& Lin 2012), where a more gradual, late evolution of $P_{\rm Sat}/P_{\rm Jup}$ would lead to an excessive dynamical excitation 
and/or collisions between the terrestrial planets. It is required from the inclination distribution of main belt asteroids 
(Morbidelli et al. 2010), where slowly sweeping secular resonances such as $\nu_{16}$ would generate a large population of 
asteroids with $i>20^\circ$, which is not observed.} 
where the orbital period ratio of Jupiter and Saturn, $P_{\rm Sat}/P_{\rm Jup}$, changes as shown in 
Figs. \ref{fig1}b and \ref{fig2}b. In the jumping-Jupiter model, the principal coupling between the inner and outer
solar system bodies occurs via secular resonances (e.g., $g_1=g_5$, $g_4=g_5$, $g=g_5$, $g=g_6$, $s=s_6$, where $g_j$, $s_j$, $s$ 
and $g$ are the fundamental precession frequencies of planetary and asteroid orbits). The secular resonances can cause 
strong orbital excitation and inhibit the accretional growth at specific radial distances from the Sun (Clement et al. 2018).  

\subsection{Interpolation of the giant planet orbits in {\tt SyMBA}} 

NM12 and Deienno et al. (2018) simulations were performed using the symplectic $N$-body codes {\tt SyMBA} (Duncan et 
al. 1998) and {\tt Mercury} (Chambers et al. 2001), respectively.
Here we want to exactly reproduce the orbital evolution of the giant planets in the selected runs and, in 
addition, include protoplanets and planetesimals in the inner solar system (Sect. 2.1). This is not easy because
the protoplanets and planetesimals need to carry mass, but if they were included as massive bodies in {\tt SyMBA} or {\tt Mercury}, 
they would affect the orbital evolution of giant planets and produce completely different orbital histories. Clement et al. 
(2018) approached this problem by performing a large suite of $N$-body simulations where the terrestrial protoplanets/planetesimals, 
giant planets and outer disk planetesimals all had mass and gravitationally interacted among themselves. This led to 
a large collection of results. The cases that at least roughly matched various 
constraints were then analyzed in more detail. This approach is not possible here where we want to be more selective (Sect 2.2). 
To deal with this issue, we develop a new method that allows us to track the giant planet orbits taken from the original 
{\tt SyMBA}/{\tt Mercury} simulations, and include terrestrial protoplanets/planetesimals that carry mass, interact between 
themselves, but do not affect the giant planets. This works as follows.  

The planetary positions and velocities in the selected instability simulations were recorded with a 1-yr cadence in Nesvorn\'y
et al. (2013) and Deienno et al. (2018). 
Our modified code {\tt iSyMBA}, where `i' stands for interpolation, then reads the planetary orbits from a file, 
and interpolates them to any required time sub-sampling (generally 0.01-0.02 yr, which is the integration time step used 
here for the terrestrial planets). The interpolation is done in Cartesian coordinates. First, the giant planets are forward 
propagated on the ideal Keplerian orbits starting from the positions and velocities recorded by {\tt SyMBA} at the beginning 
of each 1-yr interval. Second, the {\tt SyMBA} position and velocities at the end of each 1-yr interval are propagated 
backward (again on the ideal Keplerian orbits; planetary perturbations switched off). We then calculate the weighted mean of 
these two Keplerian trajectories for each planet such that progressively more (less) weight is given to the backward (forward) 
trajectory as time approaches the end of the 1-yr interval. We verified that this interpolation method produces insignificant 
errors. 

The {\tt iSyMBA} code accommodates three types of bodies: (1) interpolated giant planets, (2) fully interacting 
terrestrial protoplanets, and (3) planetesimals that carry mass, affect the terrestrial protoplanets, but do not interact among 
themselves. The original symplectic structure of {\tt SyMBA} in Poincar\'e canonical variables (heliocentric positions, 
center-of-mass linear momenta) is LD-K-D-K-LD, where LD stands for a linear drift due to the Sun's linear momentum, K for a kick 
that accounts for planetary perturbations, and D for a drift on the Keplerian orbit. This was modified to GK-LD-TK-D-TK-LD-GK,
where GK indicates the kick corresponding to the giant planet perturbations and TK is the gravitational interaction of the 
terrestrial planets and planetesimals. The tricky part in {\tt iSyMBA} is how 
to define the center of mass in different parts of the algorithm, and how to do that consistently for the terrestrial 
protoplanets/planetesimals and interpolated giant planets. See Roig et al. (2020) for a detailed description of the code.
{\tt iSyMBA} was parallelized with OpenMP directives. Through testing we established that the code optimally runs, for the 
setup described in Sect. 2.1, when $\sim$8 threads are used (performance saturates for $>10$ threads). The typical speed-up 
is $\sim3$-5.

The start of our numerical integrations with {\tt iSyMBA} (simulated time $t=0$) is assumed to coincide with $t_0$. The 
simulations are run to $t=0.8$-30 Myr (Phase1; 10 Myr in {\tt case1}, 30 Myr in {\tt case2} and 0.8 Myr in {\tt case3}). 
The instability is assumed to happen as in the original simulation, roughly at $t_{\rm inst} = 6$ Myr for {\tt case1} and 
$t_{\rm inst} = 20$ Myr for {\tt case2} (Figs. \ref{fig1} and \ref{fig2}). This is to respect the fact that the instability 
is triggered during planetary migration, and there is always certain delay between the gas disk dispersal and $t_{\rm inst}$.
In {\tt case3}, taken from Deienno et al. (2018), we test $t_{\rm inst} = 0.6$ Myr. 

The Phase-1 integrations end at $t=t_1 \simeq 1.3$-1.7 $t_{\rm inst}$. Much longer integrations are difficult to achieve with {\tt iSyMBA}, 
because the interpolation method has large requirements on the computer memory and disk storage. As the giant planets are orbitally 
decoupled by the end of Phase 1, however, the continued integrations for $t>t_1$ (Phase 2) do not need to deal with the stochastic 
outcomes of encounters between the giant planets; it therefore suffices to use the standard {\tt SyMBA} code. Phase 2 simulations are run 
to $t_2=200$ Myr, at which point the terrestrial planet accretion is complete. For the static cases {\tt nojov}, {\tt jovini} and 
{\tt jovend}, where no interpolation is needed, we use the standard {\tt SyMBA} for the whole length of simulations.

\subsection{Summary of simulations and success criteria}

Table 1 lists all simulations performed in this work. We have three static models plus three instability models, and three 
different distributions of protoplanets/planetesimals in the inner solar system. This represents $6\times3=18$ cases in total.
For each case, we perform 10 simulations where different random seeds are used to generate slightly different initial 
conditions. This represents 180 base simulations in total. In addition, we set up four additional cases that have the best 
chance to match constraints and perform 100 additional simulations for each to thoroughly characterize the statistical 
variability of the results (Sects. 3.4 and 3.5). The simulations were done on the NASA Pleiades supercomputer. Each 
simulation was completed over $\sim$3 weeks on eight Broadwell cores.
 
All collisions between protoplanets and planetesimals are assumed to end in inelastic mergers of bodies. Whether 
the debris generation in giant impacts is important for the terrestrial planet formation is still a matter of debate. 
On one hand, Mercury's mantle may have been stripped by a hit-and-run collision with a large body (Asphaug \& Reufer 2014), 
and the Moon likely formed from a circumplanetary debris disk generated by a large impact on proto-Earth (see 
Canup et al. 2020 for a review). On the other hand, Deienno et al. (2019) found no change in the terrestrial planet 
formation results when different energy dissipation schemes were used to deal with collisions. Given that the main goal of 
our work is to test the effect of giant planets on the terrestrial planet formation (and on the asteroid belt), 
we prefer to keep things simple and leave a more complex treatment of collisions for future work. The integrations reported 
here use a 5-day timestep, but we verified in several cases that the results are statistically indistinguishable 
when a 3.5-day timestep is used (the orbital period of Mercury is about 88 days, $\sim$18 and $\sim$26 times longer 
than our time steps).

Different measures are used here to to characterize the overall success of simulations. To quantify the 
radial distribution of planetary mass, Chambers (2001) defined the radial mass concentration, or RMC, as
\begin{equation}
S_{\rm c}=\max \left( {\sum_j m_j \over \sum_j m_j [\log_{10}(a/a_j)]^2} \right) \ , 
\end{equation}
where $m_j$ and $a_j$ are the mass and semimajor axis of planet $j$, the sum goes over all planets in the inner solar 
system (giant planets not included), 
and the maximum is taken over $a$; $S_{\rm c}=89.9$ in the real solar system. The AMD is defined as 
\begin{equation}
S_{\rm d}= {\sum_j m_j \sqrt{a_j} (1-\sqrt{1-e_j^2} \cos i_j ) \over \sum_j m_j \sqrt{a_j} } \ ,
\end{equation}
where $e_j$ and $i_j$ are planetary eccentricities and inclinations; $S_{\rm d}=0.0018$ in the real solar system. The 
advantage of these parameters is their objective mathematical formulation and relationship to the dynamical stability  
of planetary systems (e.g., Petit et al. 2018). Note, however, that they can be computed, and are often reported as such, for simulations 
that do not produce meaningful results (when the number and/or masses of the final terrestrial planets are clearly wrong).
This can bias the statistical interpretation of results because correlations between different measures exist (e.g., 
Deienno et al. 2019)

For these reasons, we report the $S_{\rm c}$ and $S_{\rm d}$ statistics only for the simulations that produced `good' Venus/Earth planets.
To this end, we collect all planets with mass $M > 1/3$ $M_{\rm Earth}$ between 0.5 and 1.2 au and call them the Venus/Earth analogs.
The successful simulations are required to have exactly two Venus/Earth analogs (denoted by indices 1 and 2 below; $f_{\rm g}$
in Table 1 reports the fraction of these `good' simulations for each model). We also compute the radial separation between 
good Venus/Earth planets as $\Delta a = a_2-a_1$ (this is a complementary parameter to $S_{\rm c}$).  In the real solar 
system, Venus and Earth have $\Delta a = 0.277$ au. We also determine, as a complement to $S_{\rm d}$, the mean 
eccentricities/inclinations of Venus/Earth analogs as $\langle e,i \rangle= (e_1+i_1+e_2+i_2)/4$. The real Earth and Venus 
have $\langle e,i \rangle =  0.0274$. 

The values of $\Delta a$ and $\langle e,i \rangle$ measure the radial separation and orbital excitation of the two 
large planets that form at 0.5-1.2 au. The $S_{\rm c}$ and $S_{\rm d}$ parameters are related to that but, unlike 
$\Delta a$ and $\langle e,i \rangle$, they are influenced, for example, by whether Mars/Mercury form in the simulations, and 
if so, whether they at least approximately have the right mass (e.g., massive Mars/Mercury on perfect orbits would increase 
AMD/decrease RMC not because the orbits are too excited/spread but because the planets are too massive). In this work, 
the planets with $a<0.5$ au are called Mercury and the planets with $1.2<a<1.8$ au are called Mars; they can have any mass  
$M>0.01$ $M_{\rm Earth}$ with our setup. We also define {\it strict} Mercury and {\it strict} Mars as $a<0.5$ au and $0.025<M<0.2$ 
$M_{\rm Earth}$ for Mercury, and $1.2<a<1.8$ au and $0.05<M<0.2$ $M_{\rm Earth}$ for Mars. We allow for a larger mass range in 
the case of Mercury to leave space for the possibility of hit-and-run collisions (Chambers 2013, Asphaug \& 
Reufer 2014, Clement et al. 2019b). We pay close attention to the mass of Mars planets obtained in the simulations, because small 
Mars is thought to be an important diagnostic for the early evolution of the solar system (e.g., Chambers \& Wetherill 1998, 
Chambers 2001, Hansen 2009, Walsh et al. 2011, Walsh \& Morbidelli 2011, Jacobson \& Morbidelli 2014, Clement et al. 2018).

\section{Results}

\subsection{Radial distribution of planetary mass}

\subsubsection{Model C98}

\begin{figure}
\epsscale{0.75}
\plotone{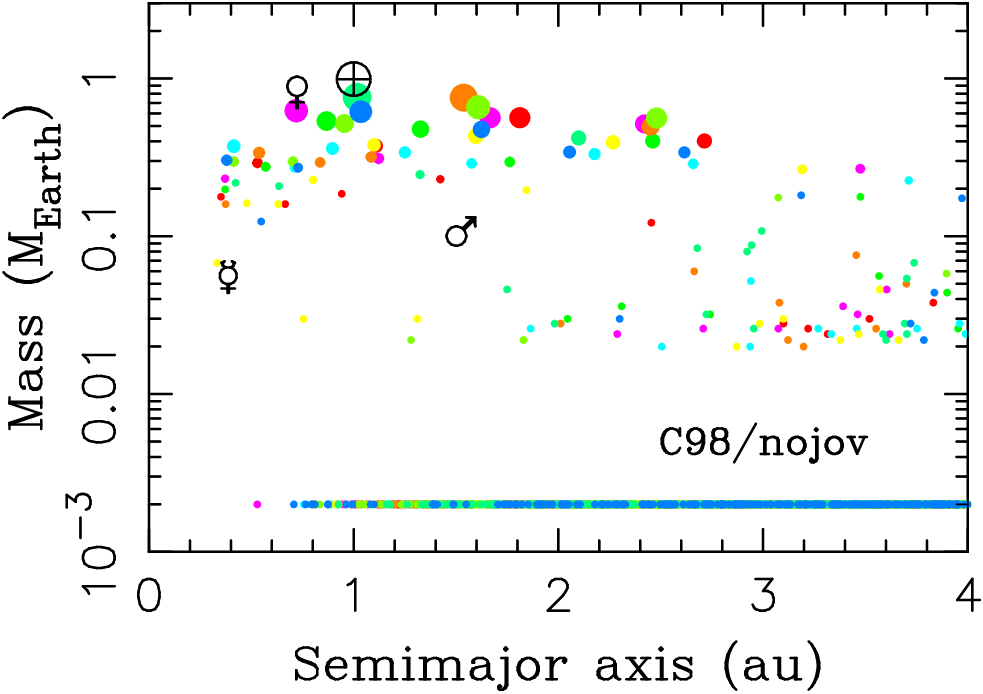}
\caption{The final mass and radial distribution of planets and planetesimals in the C98/{\tt nojov} model. The size of a dot 
increases with planet mass and its color distinguishes different simulations. The real terrestrial planets are indicated by symbols: 
$\mercury$ for Mercury, $\venus$ for Venus, $\oplus$ for Earth, $\mars$ for Mars.}
\label{s1ref}
\end{figure}

We first briefly comment on C98/{\tt nojov} to set the stage for the presentation of more successful models.
With an extended disk of protoplanets/planetesimals and no giant planets included in simulations, planets with $M>0.1$ $M_{\rm Earth}$ 
form over the entire radial range $0.3<r<4.0$ au (Fig. \ref{s1ref}). This is clearly unsatisfactory because Mars with $a=1.52$ au is the 
outermost terrestrial planet in the real solar system, and there are no planetary-size bodies in the asteroid belt ($2.2<r<3.5$ au).
The masses of Venus/Earth analogs are somewhat small indicating that the total mass in solids placed into the inner solar system, 
$M_{\rm solid}=4$ $M_{\rm Earth}$, was not large enough in this model. We increase the total mass of protoplanets/planetesimals in the 
models described in Sect. 3.4. Also, the planets that form in C98/{\tt nojov} at $r \sim 1.5$ au are excessively massive, 
typically $\sim$2-7 times more massive than the real Mars. 

\begin{figure}
\epsscale{0.75}
\plotone{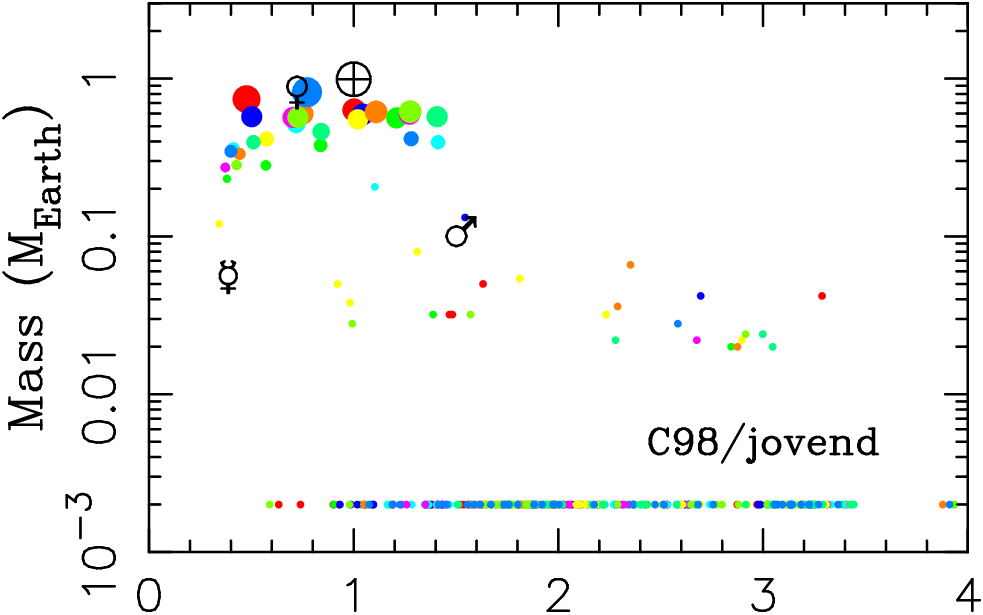}\\[2.mm]
\plotone{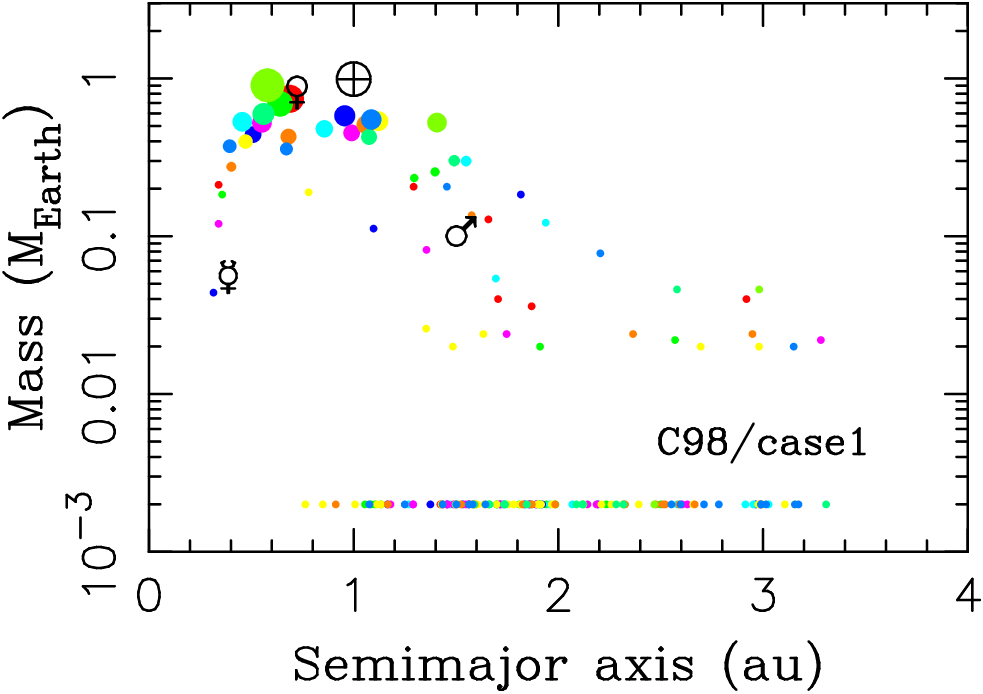}
\caption{The final mass and radial distribution of planets and planetesimals in the C98/{\tt jovend} (top panel) and 
C98/{\tt case1} (bottom panel) models. See caption of Fig. \ref{s1ref} for the meaning of different symbols.}   
\label{s1}
\end{figure}

The results improve when the giant planets are included in the simulations. With the giant planets on their 
current orbits (model C98/{\tt jovend}; Fig. \ref{s1}),\footnote{If the giant planets are placed on their pre-instability 
resonant orbits in C98/{\tt jovini}, they become unstable at $\sim30$-100 Myr after the start of simulations. This happens 
because the giant planets gravitationally interact with the protoplanets that start at $2<r<4$ au and evolve onto Jupiter-crossing orbits. 
Recall that the massive outer disk at $r>20$ au is not included in our simulations. The giant planet instabilities become violent 
and these cases do not represent plausible evolutionary histories of the early solar system. The instability does not happen in the 
W11/{\tt jovini} and L15/{\tt jovini} models where no protoplanets start at $r>1.5$ au.} there is a sharp transition from $r<1.5$ au, 
where massive planets with $M>0.2$ $M_{\rm Earth}$ grow, to $r>1.5$ au, where the growth is modest and the final planets have masses
$0.02<M<0.2$ $M_{\rm Earth}$. A similar result has been reported in Raymond et al. (2009; their case EJS). The transition is most 
likely related to the combined effect of protoplanet scattering and secular resonances $\nu_5$, $\nu_6$ and $\nu_{16}$ that excite 
orbits near $a = 2$ au. Beyond 2 au, protoplanet scattering, orbital resonances with Jupiter and the proximity to Jupiter eliminate 
most protoplanets and planetesimals from the asteroid belt region ($f_{\rm ast} \simeq 0.02$; Table 1).
   
In the C98/{\tt jovend} model, the transition at $r=1.5$ au is too sharp with planets near 1.5 au either having $M > 0.3$ 
$M_{\rm Earth}$ or $M < 0.05$ $M_{\rm Earth}$. This leaves Mars in the no man's land. In fact, there are only two relatively good 
Mars planets (small yellow and blue dots in the top panel of Fig. \ref{s1}) in ten simulations, which is not very satisfactory. 
The results are better when the giant planet instability is taken into account (C98/{\tt case1} in the bottom panel of Fig. \ref{s1}). 
In this case, the planetary mass shows a more gradual decrease with the radial distance: from $M \gtrsim 0.5$ $M_{\rm Earth}$ at 
$r=1$ au, to $M \sim 0.1$-0.3 $M_{\rm Earth}$ at $r=1.5$ au, to $M \sim 0.02$-0.1 $M_{\rm Earth}$ at $r=2$ au. The mass scatter, 
however, is quite large and only two of our ten C98/{\tt case1} simulations produced acceptable matches to the inner solar 
system (two planets with $M>1/3$ $M_{\rm Earth}$ and $0.5<a<1.2$ au, $\Delta a < 0.415$ au -- no more than 1.5 times the real 
value -- and $\langle e,i \rangle < 0.0548$ -- less than a double of the real value -- and the simulated Mars mass within a 
factor of two to the real mass). The results for C98/{\tt case3} are similar (Table 1). We therefore do not see much dependence on 
the orbital evolution of giant planets during the instability (at least for the class of instabilities investigated 
here; recall that the {\tt case1} and {\tt case3} instabilities are quite different, Sect 2.2). Interestingly, the C98/{\tt case2} 
model does not produce any strict Mars analogs (planets near 1.5 au are too massive; Table~1). This is probably a consequence
of the fact that the giant planet instability happens relatively late in {\tt case2} ($t_{\rm inst} \simeq 20$ Myr), and the 
planets near $r=1.5$ au accrete too much mass before this region is affected by the giant planets (Clement et al. 2018).

In summary, the primary effect on the terrestrial planet formation is the presence of the giant planets themselves (if they reach their
current orbits early) that creates the transition at 1.5 au. The secondary effect is the evolving orbits of the giant planets during the instability, 
which makes the transition more gradual and allows for the formation of small (but not too small) Mars at 1.5 au. For that, however, 
the giant planet instability must happen relatively early (before $10$ Myr after $t_0$ or so; Clement et al. 2018). If it would happen 
late ($t>10$ Myr), the Mars's growth would not be affected by it.

Interestingly, there are some reasonable Mercury analogs in the C98/{\tt case1} model, where four out of our ten simulations produced 
planets with $M \lesssim 0.2$ $M_{\rm Earth}$ and $a \simeq 0.4$ au (bottom panel of Fig. \ref{s1}). Only one such planet was obtained in ten
simulations of the C98/{\tt jovend} model. This indicates that having small Mercuries in C98/{\tt case1} is probably not a consequence 
of the inner edge of the initial disk (set at $r_{\rm in}=0.3$ au to limit the CPU cost). The orbits of Mercury analogs are excited to 
eccentricities $e\simeq0$-0.25 and inclinations $i\simeq5$-15$^\circ$. This is reminiscent to the results of Roig et al. (2016), who 
showed that the orbit of fully formed Mercury could be excited by sweeping secular resonances ($g_1=g_5$ and $s_1=s_7$), 
if the giant planet migration/instability happened relatively late. We therefore believe that the same resonances act here, 
during the terrestrial planet formation, to excite orbits at $r \simeq 0.4$ au and stall Mercury's growth.  
 
We find that $N_{\rm spl}=1.5$  and 0.4 planetary-mass bodies ($M>0.01$ $M_{\rm Earth}$, average per good simulation, Table 1, index 
`spl' stands for `stranded planets') end up in the asteroid belt in the C98/{\tt jovend} and C98/{\tt case1} models, respectively. 
These results are therefore more plausible than C98/{\tt nojov} where $N_{\rm spl}=4.5$. Planetary-mass bodies in the asteroid belt 
would have a huge influence on the depletion, mixing and orbital excitation of asteroids (Sect. 3.2). The fact that $N_{\rm spl}\sim 1$ 
in the C98/{\tt jovend} and C98/{\tt case1} models is still troublesome but this can potentially be resolved if: (i) the long-term 
instabilities remove additional bodies from the asteroid belt on Gyr timescales, and/or (ii) the original protoplanet/planetesimal 
disk in the inner solar system had a steep radial profile (Levison et al. 2015, 
Izidoro et al. 2015). Here we use $\gamma=1$ (Sect. 2.1). Simulations 
with $\gamma=1.5$ are described in Section 3.4. We proceed by discussing the results of the W11 and L15 setups where $N_{\rm spl} \sim 0$ 
(except if the {\tt nojov} or {\tt jovini} models are considered; Table 1).

\subsubsection{Models W11 and L15}

\begin{figure}
\epsscale{0.75}
\plotone{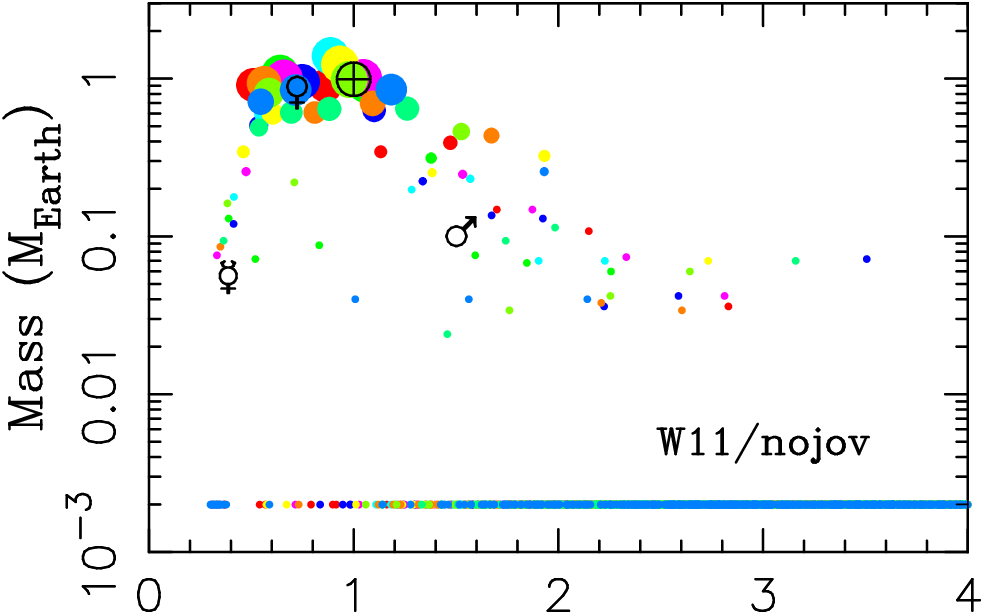}\\[2.mm]
\plotone{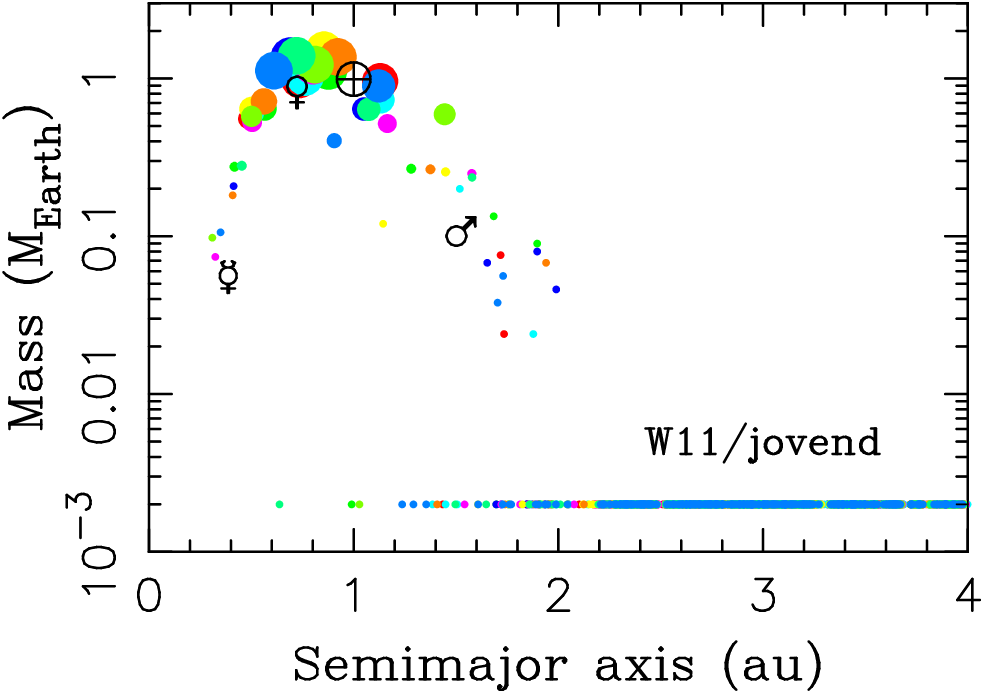}
\caption{The final mass and radial distribution of protoplanets and planetesimals in the W11/{\tt nojov} (top panel) and 
W11/{\tt jovend} (bottom panel) models. See caption of Fig. \ref{s1ref} for the meaning of different symbols.}
\label{s2}
\end{figure}

Figure \ref{s2} compares the results for the W11/{\tt nojov} and W11/{\tt jovend} models. The protoplanets with $M=0.02$ $M_{\rm Earth}$ 
start in a narrow annulus in these models (Hansen 2009, Walsh et al. 2011, Walsh \& Levison 2016), 
and the radial mass profile of planets at $t=200$ Myr is 
peaked near the original annulus location. The radial distribution of final planets ($0.5<r<1.2$ au at $t=200$ Myr), however, 
is wider than that of the original annulus ($0.7<r<1$ au at $t_0$). This happens because there is no active mechanism that would confine 
protoplanets to the annulus. Instead, as protoplanets grow, they scatter each other, and the general tendency of this 
relaxation process is to spread bodies over a larger range of radial distances. An undesired consequence 
of this process in that our simulations typically lead to a relatively large separation between Venus and Earth (Sect. 3.3.1). 
This problem could potentially be resolved if the initial protoplanets were more massive (Jacobson \& Morbidelli 2014). We test
this possibility in Sect. 3.5.

The effect of giant planets on the terrestrial planet formation is not as dramatic in the narrow annulus model (Fig. \ref{s2}) 
as it is in the C98 model (Figs. \ref{s1ref} and \ref{s1}). There are far fewer planets stranded in the asteroid belt 
in W11/{\tt nojov} ($N_{\rm spl}=1.4$) than in C98/{\tt nojov} ($N_{\rm spl}=4.5$), and including the giant planets in W11 reduces 
$N_{\rm spl}$ to zero (except for {\tt jovini} where $N_{\rm spl}=0.6$). The results for W11/{\tt jovend} (bottom panel of Fig. \ref{s2}) 
are relatively good. The Mars planets at 1.2-1.8 au have $M=0.02$-0.3 $M_{\rm Earth}$, except for one case (large green dot) where the
massive planet with $M\simeq 0.6$ $M_{\rm Earth}$ formed near 1.42 au. 

Model Mercuries with $M<0.2$ $M_{\rm Earth}$ and 
$0.3<r<0.5$ au form in both W11/{\tt nojov} and W11/{\tt jovend}. In this case, the primary reason for a small planet at $r\sim0.4$ 
au is the narrow annulus setup and, consequently, the lack of protoplanets initially 
available for accretion at $r<0.7$ au. Model Mercuries in W11 are 
protoplanets scattered from $r>0.7$ au to $r \sim 0.4$ au, where their orbits become stabilized by dynamical friction from 
planetesimals. The model planets at $r\sim0.4$ au are often a factor of $\sim$1.5-4 more massive than Mercury, but this may
not be an insurmountable problem because Mercury's mass could have been reduced by violent hit-and-run collisions 
(Chambers 2013, Asphaug \& Reufer 2014, Clement et al. 2019b), a process not accounted for here.

\begin{figure}
\epsscale{0.75}
\plotone{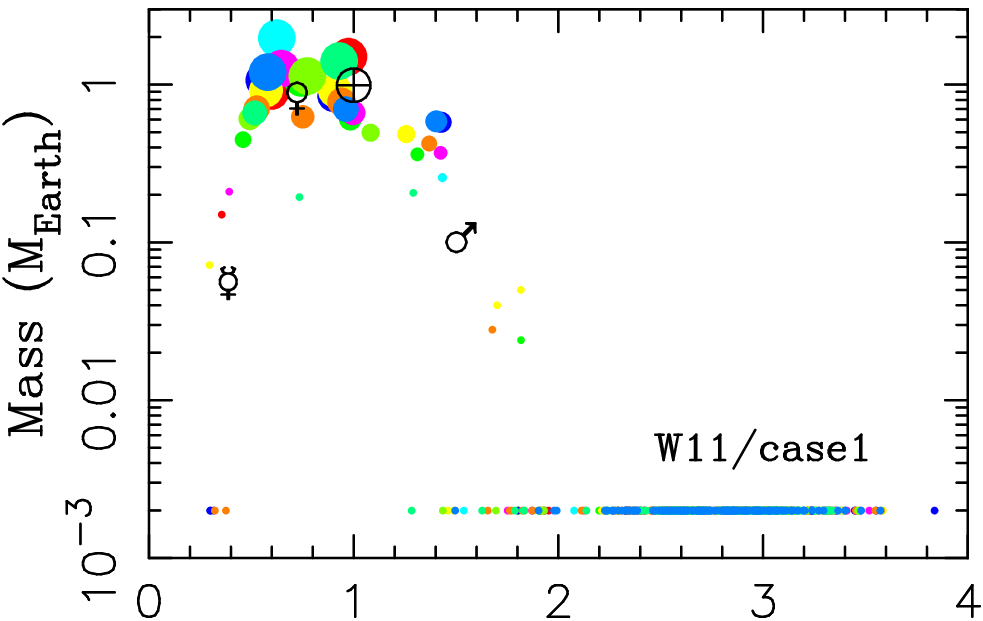}\\[2.mm]
\plotone{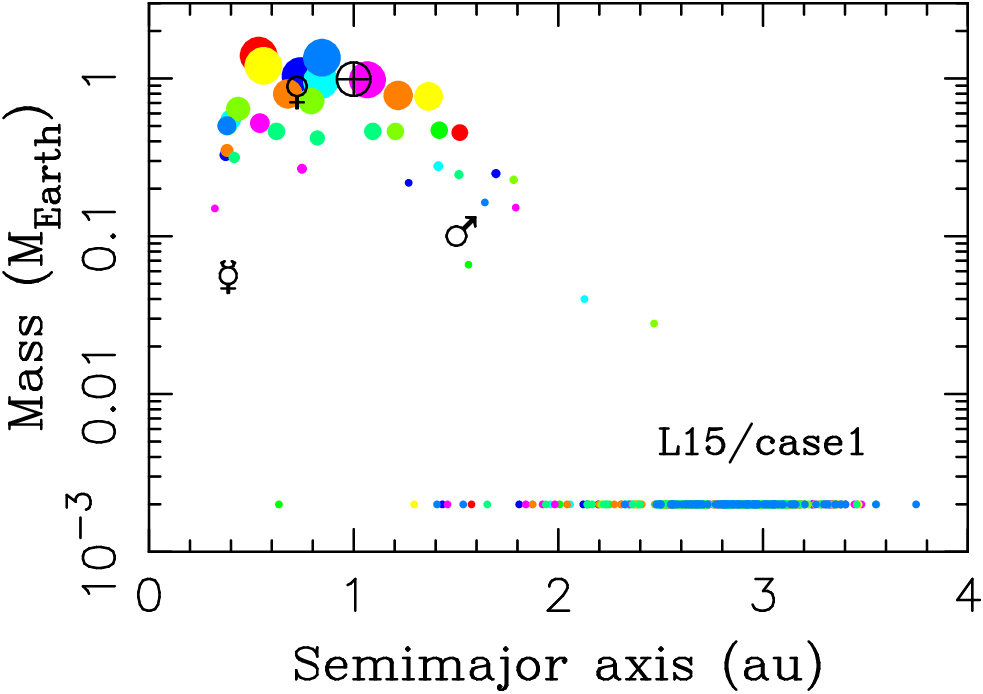}
\caption{The mass and radial distribution of planets and planetesimals in the W11/{\tt case1} (top panel) and 
L15/{\tt case1} (bottom panel) models. See caption of Fig. \ref{s1ref} for the meaning of different symbols.}
\label{case1}
\end{figure}

The results with and without the instability, W11/{\tt case1} in the top panel of Fig. \ref{case1} and 
W11/{\tt jovend} in the bottom panel of Fig. \ref{s2}, 
are similar in that they show a relatively strong concentration of massive planets near the original 
annulus location. There are interesting differences as well. Notably, the W11/{\tt case1} model produced fewer Mars/Mercury 
planets than the W11/{\tt jovend} model. This is probably a consequence of the coupled effect of the narrow annulus and 
sweeping secular resonances in W11/{\tt case1}. The results for W11/{\tt case3} show the same trend but the statistics of 
Mars/Mercury planets is better in the W11/{\tt case2} model. This probably indicates some mild dependence on the instability case.
 
The L15/{\tt case1} model (bottom panel of Fig. \ref{case1}) does not offer any obvious advantage over the W11/{\tt case1} model. 
The final radial mass distribution of planets in L15/{\tt case1} is relatively broad, probably because the original protoplanet disk 
has an edge at $r_{\rm out}=1.5$ au in L15, and there are fewer Mercuries than in W11, probably because there are no planetesimals 
(i.e., no dynamical friction) to stabilize protoplanets scattered from the original annulus location to $r\sim0.4$ au. We 
see no important differences between the results obtained for different instability cases and thus, for the sake of brevity, 
we do not discuss the {\tt case2} and {\tt case3} models in detail here (see Sect. 3.3 for a more detailed comparison). 

\subsubsection{Comparison of the C98, W11 and L15 models}

We could adopt a viewpoint that it is not known whether the terrestrial planets formed from a narrow annulus or some other radial 
distribution of solids, but we are reasonably sure that the giant planet migration/instability happened (given the large 
number of constraints that we have; Nesvorn\'y 2018). It would then 
be appropriate to compare the results obtained for different radial profiles (C98, W11, L15) and the same migration/instability 
case (e.g., {\tt case1}) to see if there are any important differences. 
This can be done by inspecting Figs. \ref{s1} (bottom panel for C98/{\tt case1}) and \ref{case1} 
(top for W11/{\tt case1} and bottom for L15/{\tt case1}). There are two main trends. The first one was already mentioned: 
as the initial radial distribution becomes more concentrated toward $r=0.7$-1 au (progression C98 $\rightarrow$ L15 $\rightarrow$ W11), 
the final radial distribution of planets becomes more concentrated near $r=0.7$-1 au as well. For this reason, given the tight 
orbital separation of Venus and Earth, a preference should be given to the W11/{\tt case1} model. The second and related trend is 
that the same progression generates fewer protoplanets in the asteroid belt region, which is good, and fewer Mars analogs, 
which is bad. This leads to an ugly conclusion that the narrow annulus and giant planet effects do not seem to mash 
constructively (at least for the cases studied here).

\subsection{Orbital excitation}

The orbits of the terrestrial planets, and their orbital eccentricities and inclinations in particular, are important constraints
on the planet/planetesimal formation in the inner solar system. These constraints are best evaluated together with the dynamical 
structure of asteroid belt, because: (i) the main belt asteroids are the only leftover planetesimals that survived in the inner 
solar system, and (ii) their overall mass and orbital distribution have been influenced both by protoplanet formation processes 
and the giant planet instability. 

\begin{figure}
\epsscale{0.65}
\plotone{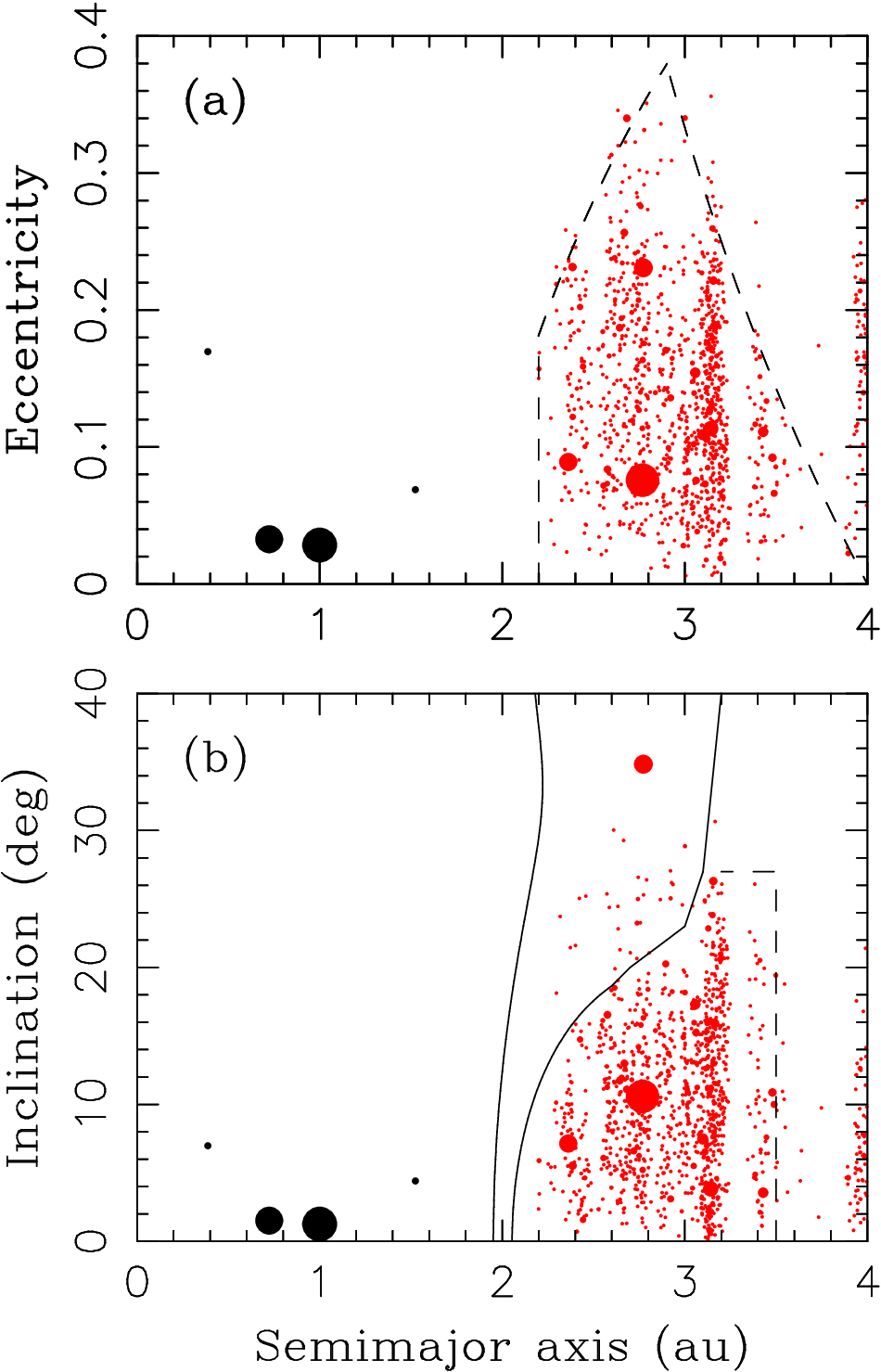}
\caption{The mean orbital elements of the terrestrial planets (black dots) and asteroids (red dots). The size of 
a black dot is proportional to the planet mass, and the size of a red dot is proportional to the asteroid size. 
Only asteroids with diameters $D>30$ km are plotted here. The lines in panels (a) and (b) approximately carve out 
the asteroid belt region in $(a,e)$ and $(a,i)$. The dashed lines correspond to $a=2.2$ au, $q=a(1-e)=1.8$ au and 
$Q=a(1+e)=4$ au (panel a), and $a=3.5$ au and $i=25^\circ$ (panel b). The solid lines in (b) are the secular 
resonances $\nu_6$ and $\nu_{16}$ (plotted here for $e \sim 0$).}
\label{real}
\end{figure}

Figure \ref{real} shows the real orbits of four terrestrial planets and all known asteroids with diameters $D>30$ km. We only plot the
large asteroids because they are less affected by radiation effects and collisions than the small ones; their orbital
distribution is thus more diagnostic for processes in the early solar system. Mercury's relatively large orbital eccentricity 
(mean $e \simeq 0.17$) and inclination (mean $i \simeq 7^\circ$) is a puzzling aspect of the inner solar system. In the narrow 
annulus model, proto-Mercury is scattered from $r=0.7$-1 au and stabilized at its current orbit with $a \simeq 0.38$ au 
by collisions and/or dynamical friction. In Roig at al. (2016), Mercury can start with $e \sim i \sim 0$ and its orbit becomes 
excited during the giant planet instability.

The main asteroid belt spans $r \simeq 2.2$-3.5 au (here we ignore Hildas in the 3:2 resonance with Jupiter, $a\simeq4$ au;
see Roig \& Nesvorn\'y 2015). The orbital eccentricities of asteroids are relatively large -- the orbits practically fill the whole 
region that is dynamically stable. The eccentricities could have been low at $t_0$ and increased after $t_0$, for example, 
during the instability (Deienno et al. 2018, Clement et al. 2019a) and/or by protoplanet scattering (O'Brien et al. 2007, 
Clement et al. 2019a). 
The eccentricity distribution could have also been quite broad at $t_0$, as in the Grand Tack model (Walsh et al. 2011; also see 
Izidoro et al. 2016), and evolved to a narrower distribution as the high-$e$ orbits were removed over 4.5~Gyr (Roig 
\& Nesvorn\'y 2015, Deienno et al. 2016). 

The inclination distribution 
of asteroids represents a more meaningful constraint. The high-$i$ orbits with $i\simeq20$-30$^\circ$ are generally stable 
and yet the number of asteroids falls off for $i>20^\circ$ (Fig. \ref{real}b). This means that a relatively small population 
of asteroids had $i\simeq20$-30$^\circ$ in the first place, which is an interesting constraint on the Grand Tack model (Deienno et 
al. 2016) and dynamical evolution of the giant planets after $t_0$ (Morbidelli et al. 2010, Roig \& Nesvorn\'y 2015, 
Clement et al. 2020a).

\begin{figure}
\epsscale{0.65}
\plotone{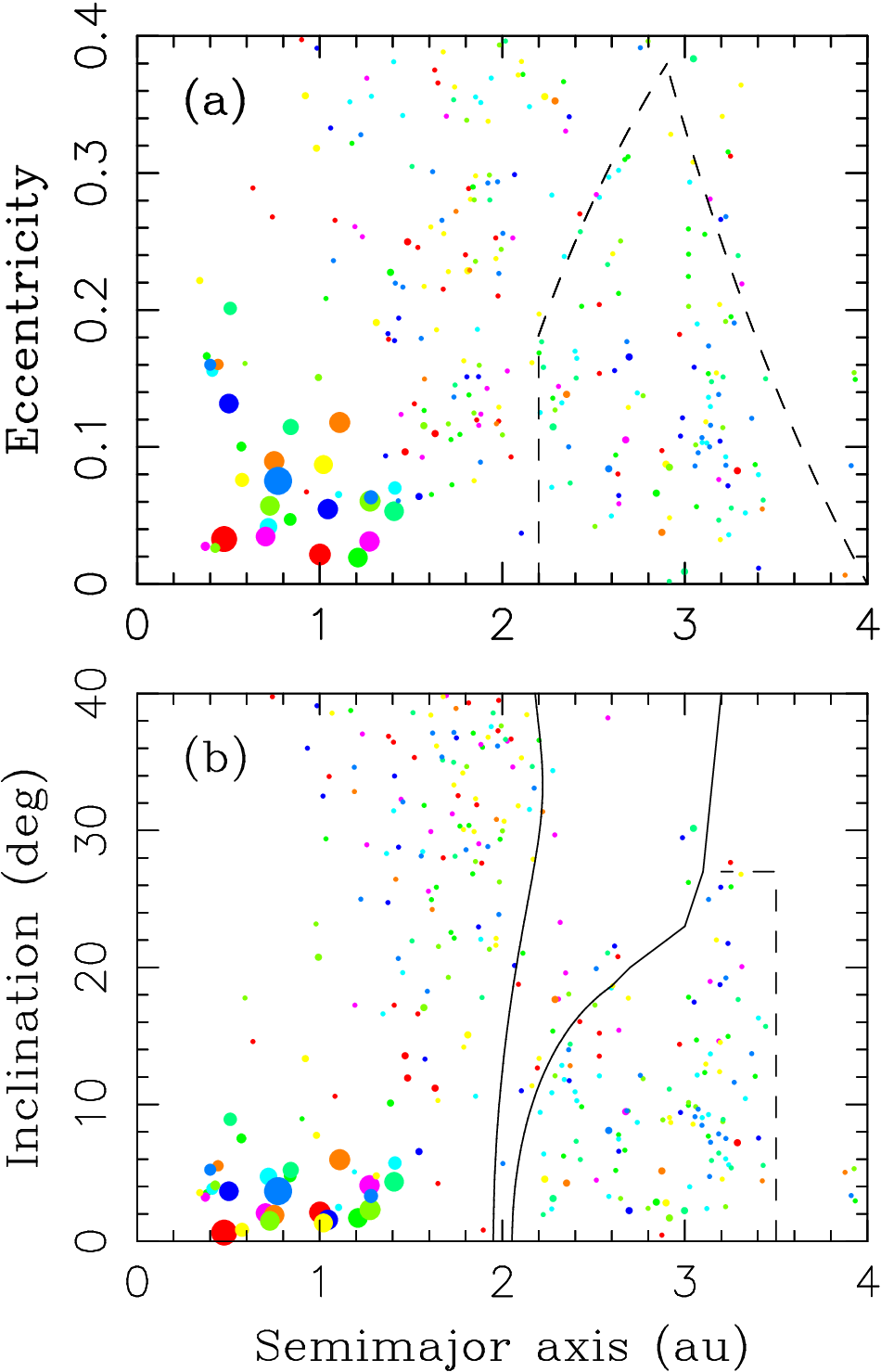}
\caption{The orbits of planets and asteroids obtained in the C98/{\tt jovend} model. See caption of
Fig. \ref{real} for the meaning of dots and lines. The color highlights the results obtained in ten different
simulations of C98/{\tt jovend}.}
\label{s1orb}
\end{figure}

\begin{figure}
\epsscale{0.65}
\plotone{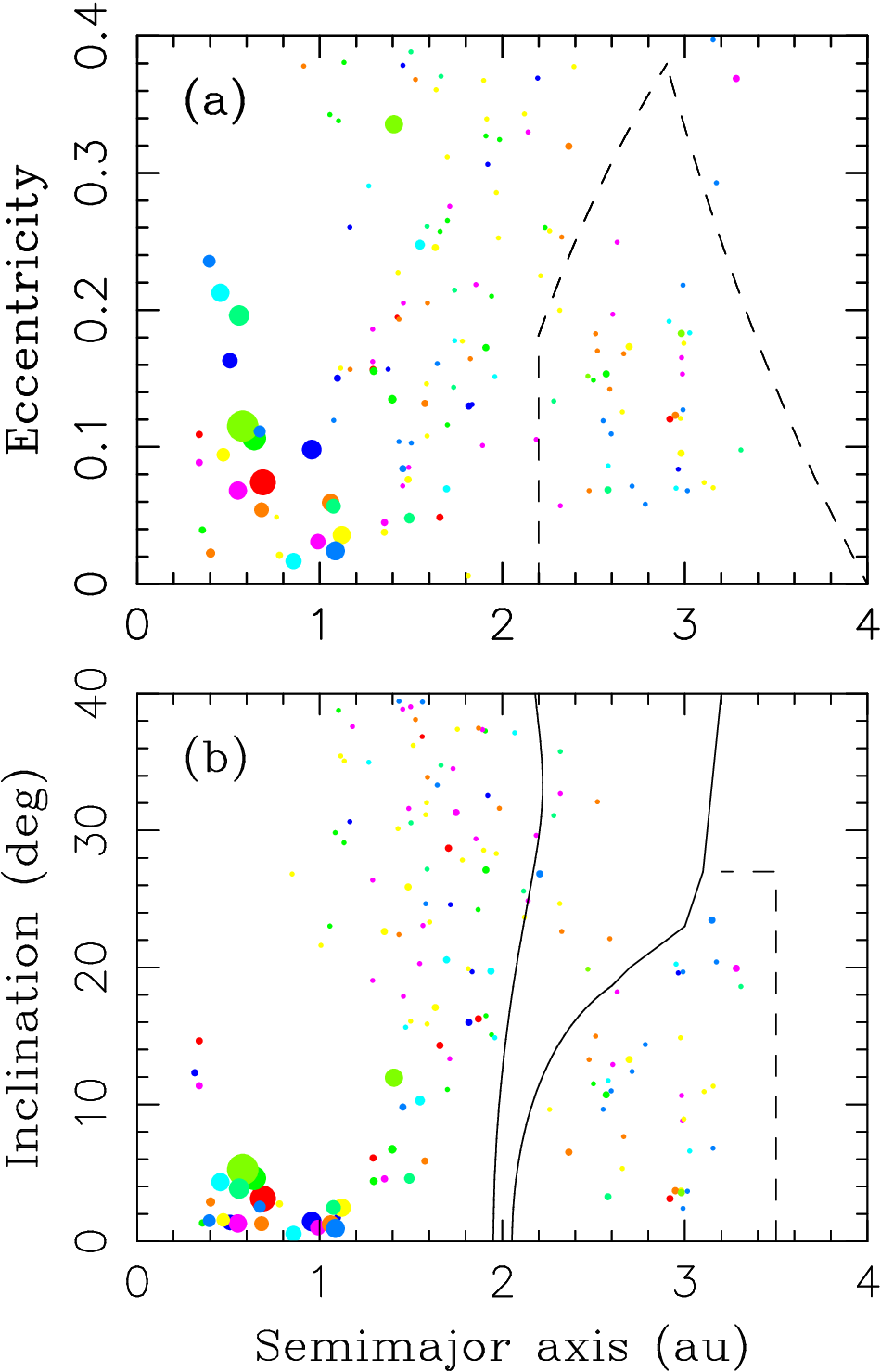}
\caption{The orbits of planets and asteroids obtained in the C98/{\tt case1} model. See caption of
Fig. \ref{real} for the meaning of dots and lines. The color highlights the results obtained in ten different
simulations of C98/{\tt case1}.}
\label{s1orb2}
\end{figure}

Figures \ref{s1orb} and \ref{s1orb2} show the orbits of planets and asteroids for the C98 model. Both in the {\tt jovend}
and {\tt case1} models the small planets at $r<0.5$ au and $r \sim 1.5$ au tend to have more excited orbits than the massive 
planets at $r \simeq 0.5$-1.2 au. Obtaining good orbits for Mercury and Mars may thus not be difficult 
(we would need a larger statistic to be able to evaluate this in detail).
Instead, the main problem is that the orbital eccentricities and inclinations of massive planets at 0.5-1.2 au 
are generally too large (Sect 3.3). This problem -- traditionally expressed in terms of the total AMD of the terrestrial planets -- 
received much attention since it was first pointed out in Chambers \& Wetherill (1998). Increasing the planetesimal
population would help because this enhances the effects of dynamical friction on protoplanets (e.g., O'Brien et al. 2006),
but Jacobson \& Morbidelli (2014) and Levison et al. (2015) argued for a relatively small mass in planetesimals at 
$r \sim 1$ au. Here we split the initial mass equally in planets and planetesimals.

As for the orbital structure of the asteroid belt, we find that the asteroid orbits are strongly but not excessively 
excited in the C98 model. The 
planetesimal population at $r=2.2$-3.5 au is depleted by factor of $\sim 50$ in C98/{\tt jovend} and $\sim 100$ in 
C98/{\tt case1} (Table 1), and there are on average 1.5 and 0.4 planets stranded in the asteroid belt at $t=200$ Myr
in C98/{\tt jovend} and C98/{\tt case1}, respectively. In these models, the excitation and depletion of asteroid 
orbits is mainly caused by the protoplanets that start in the main belt at $t_0$. Given that the number of protoplanets 
surviving in the main belt at $t=200$ Myr is generally low (some would be removed during the subsequent evolution,
a steeper radial profile would limit their number as well; Sect. 3.4 and O'Brien et al. 2007), the C98 model cannot 
be ruled out based on this criterion alone. The maximum depletion factor $\sim140$ is obtained for C98/{\tt case2}.
Interestingly, at $t=200$ Myr, many leftover planetesimals have planet-crossing orbits with $a=1$-2 au and orbital 
inclinations reaching up to $i \simeq 40^\circ$ (Figs. \ref{s1orb} and \ref{s1orb2}). The potential relevance of this 
impactor population for the Late Heavy Bombardment (LHB) will be addressed in our future work.

\begin{figure}
\epsscale{0.65}
\plotone{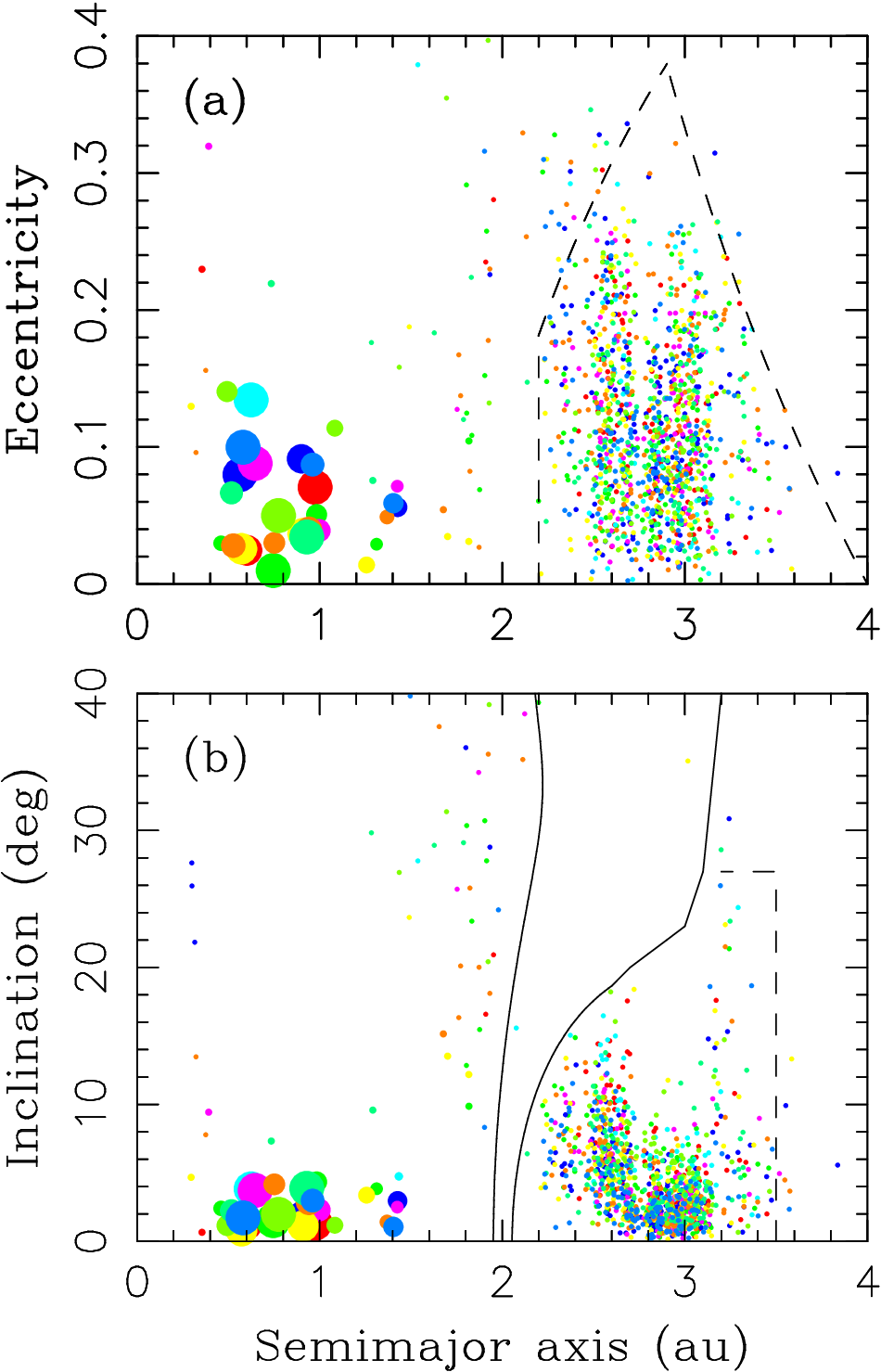}
\caption{The orbits of planets and asteroids obtained in the W11/{\tt case1} model. See caption of
Fig. \ref{real} for the meaning of dots and lines. The color highlights the results obtained in ten different
simulations of W11/{\tt case1}.}
\label{s2orb}
\end{figure}

\begin{figure}
\epsscale{0.65}
\plotone{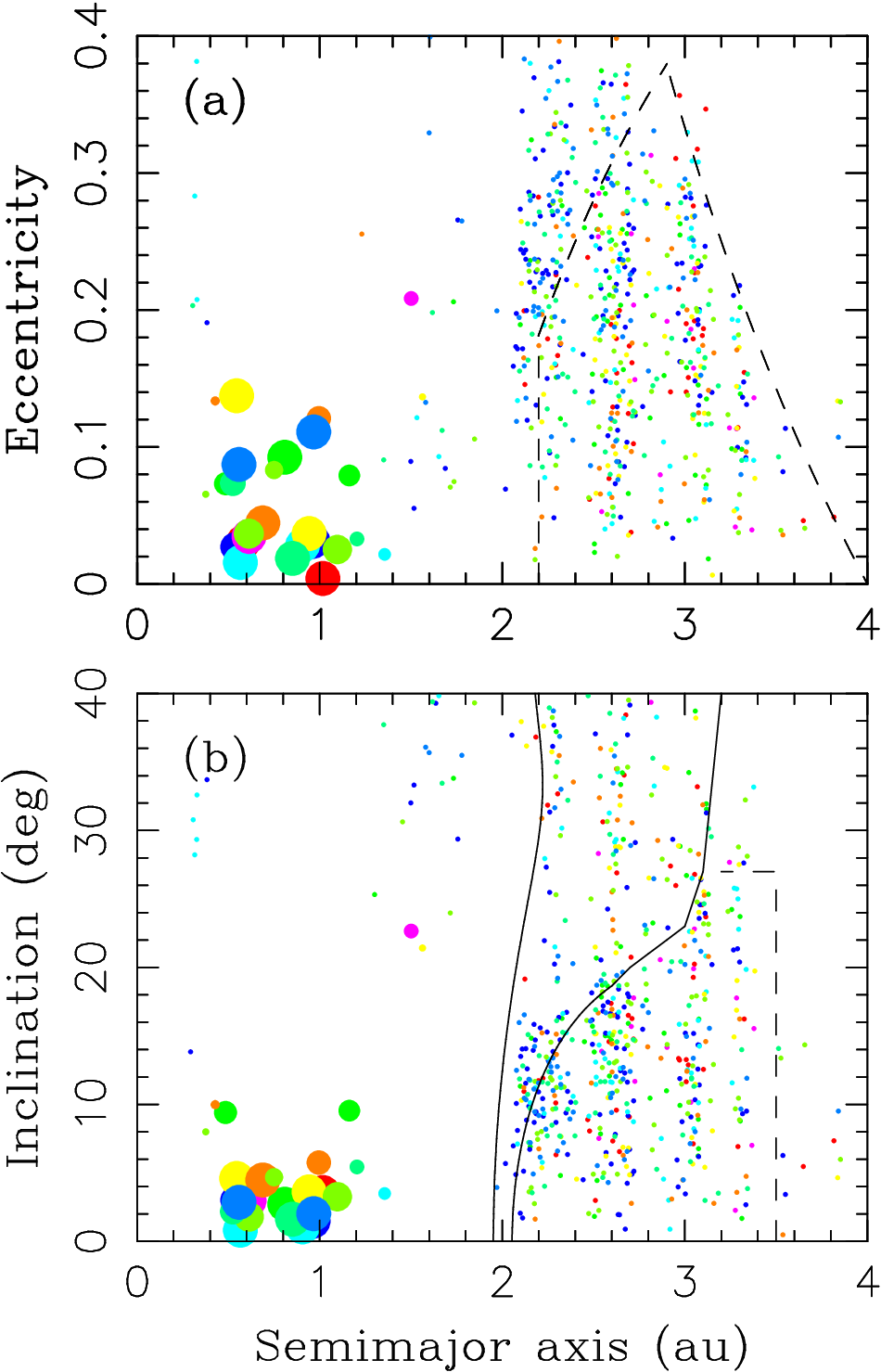}
\caption{The orbits of planets and asteroids obtained in the W11/{\tt case3} model. See caption of
Fig. \ref{real} for the meaning of dots and lines. The color highlights the results obtained in ten different
simulations of W11/{\tt case3}.}
\label{s2orb2}
\end{figure}

The orbital structure of the terrestrial planet system in the W11 model is similar (Figs. \ref{s2orb} and \ref{s2orb2}) 
with the eccentricities and inclinations of Venus/Earth planets being again somewhat too large on average. In this case,
there are no protoplanets in the asteroid belt and the excitation of asteroid orbits is therefore due to the giant
planet instability itself and, to a lesser degree, due to gravitational scattering from protoplanets whose orbits temporarily 
overlap with the main belt. The excitation of asteroid inclinations ends up to be inadequate in W11/{\tt case1} (Fig. 
\ref{s2orb}b), which is consistent with the results of Roig \& Nesvorn\'y (2015) who used the same instability 
model. In the W11/{\tt case1} model, some process would have to excite asteroid orbits before $t_0$ (e.g., Walsh et 
al. 2011). Alternatively, for the excitation to happen after $t_0$, a stronger instability model would have to be adopted   
(Deienno et al. 2018, Clement et al. 2018). Here we illustrate this for W11/{\rm case3} (Fig. \ref{s2orb2}), where 
{\tt case3} was taken from Deienno et al. (2018). The results obtained for the L15 model are very similar 
to those obtained for W11, both in what concerns the excitation of the terrestrial planet orbits, and the depletion 
and excitation of the asteroid belt (Table 1).

\subsection{Overall success of different models to match constraints}

There is a substantial freedom in how to communicate the results of the terrestrial planet simulations. For example, the 
$S_{\rm c}$ and $S_{\rm d}$ parameters can be reported for a simulation independently of whether the simulations produced 
one, two or many terrestrial planets. While this is still useful to demonstrate different trends, it is not ideal to clump 
the results with different numbers of the terrestrial planets together because various correlations exist. For example, Deienno 
et al. (2019) pointed out that $S_{\rm d}$ is correlated with the final number of planets and the systems with more planets 
typically have lower $S_{\rm d}$, and therefore appear to be more plausible in terms of $S_{\rm d}$. But, strictly speaking, 
the results with more than four terrestrial planets are incorrect and they should not be included in the $S_{\rm d}$ 
(and $S_{\rm c }$) statistics in the first place. 

As Venus and Earth are the chief contributors to $S_{\rm c}$ and $S_{\rm d}$ in the real solar system, we find it useful to 
identify simulations where some sort of reasonable Venus/Earth analogs form, and report the results only for those 
simulations. Here we collect simulated planets with $0.5<a<1.2$ au and $M>1/3$ $M_{\rm Earth}$ (Sect. 2.4), call them 
Venus/Earth, and only consider the runs where the final number of Venus/Earth planets is equal to two. The parameter
$f_{\rm g}$ (second column in Table 1), where index `g' stands for `good', reports the fraction of simulations in each 
case that produced exactly two Venus/Earth-analog planets. For example, in the W11 model, $f_{\rm g}=0.5$-0.8, indicating 
that a satisfactorily large fraction of simulations ended up with two large planets at 0.5-1.2 au. 

\subsubsection{Radial mass concentration}

All results reported in Table 1 show RMC values that are significantly lower than the actual $S_{\rm c}=89.9$, thus indicating 
that the simulated systems are typically less tightly packed than the real terrestrial planets. This is especially
obvious for C98 and L15, where $S_{\rm c}=22$-37, some 2.4 to 4 times lower than for the real planets. The giant planet 
instability does not help much to increase the RMC values (other than producing smaller Mars, but this does not project into
the $\Delta a$ parameter -- see below). This finding is similar to the results of Clement et al. (2018) who also found 
that the RMC values are generally inadequate if the protoplanet disk initially extends beyond $\sim 1$ au. The results are better 
in the W11 model, where $S_{\rm c}=40$-61. This is consistent with the results of Jacobson \& Morbidelli (2014). To obtain 
$S_{\rm c} \sim 90$, Jacobson \& Morbidelli (2014) further suggested that the late stage of the terrestrial planet formation started 
with Mars-mass protoplanets at $t_0$. If the planets are initially large, less radial spreading occurs and this leads to better 
results (Sect. 3.4; also see Lykawka \& Ito 2019, Lykawka 2020).

\subsubsection{Angular momentum deficit}

The AMD values reported in Table 1 are generally larger than $S_{\rm d}=0.0018$ of the real terrestrial planets, but there
are a few cases where the simulated AMD is much lower ($S_{\rm d}=0.0002$-0.0008; e.g., {\tt nojov} and {\tt jovini} in W11 
and L15). This happens because the Venus/Earth planets that form in these models are too small ($M \simeq 0.3$-0.5 
$M_{\rm Earth}$), their gravitational interaction is weak, and they therefore end up on orbits with a very low AMD. It is notable 
that the very-low-AMD cases are obtained in the models where there are no giant planets or the giant planet 
are placed on their pre-instability orbits (resonant with low-$e$/$i$).  
A larger mass in protoplanets would need to be initially placed at $\sim$ 1 au for planets to grow larger in these models.
The same issue affects, to a lesser degree, other models where $S_{\rm d} \lesssim 0.002$ (e.g., C98/{\tt nojov}, C98/{\tt case2}), 
except for W11/{\tt jovend} and L15/{\tt case1} where the Venus/Earth analogs have the right mass. The planetary masses and 
AMD values in the W11/{\tt case1} and W11/{\tt case3} models are relatively good (average $S_{\rm d}=0.0029$ and $S_{\rm d}=0.0024$). 
We conclude that obtaining the correct AMD for the terrestrial planets appears to be less problematic than obtaining the correct RMC 
(Deienno et al. 2019, Lykawka \& Ito 2019, Lykawka 2020).

\subsubsection{$\Delta a$ and $\langle e,i \rangle$}

The $\Delta a$ and $\langle e,i \rangle$ values reported in Table 1 are generally correlated with $S_{\rm c}$ 
and $S_{\rm d}$, but there are several interesting differences as well. Recall that $\Delta a$ and $\langle e,i \rangle$ are 
defined only from the orbital properties of Venus and Earth, but $S_{\rm c}$ and $S_{\rm d}$ are based on all terrestrial planets 
(here for systems with two Venus/Earth planets). So, for example, whereas W11/{\tt case3} shows the luringly good values of 
$S_{\rm c}=60.9$ and $S_{\rm d}=0.0024$, this happens because strict Mars/Mercury analogs often do {\it not} form in this model,
and this biases the RMC and AMD to higher and lower values, respectively (relatively to other models where massive Mercury/Mars 
planets form on excited orbits). If the $\Delta a$ and $\langle e,i \rangle$ parameters are considered instead, W11/{\tt case3} 
is not exceptionally good. Some of the best results in terms of $\Delta a$ and $\langle e,i \rangle$ are obtained for 
W11/{\tt jovend} ($\Delta a=0.35$ au, $\langle e,i \rangle=0.041$) and W11/{\tt case1} ($\Delta a=0.34$ au, $\langle e,i 
\rangle=0.048$). This is to be compared to the real $\Delta a=0.28$ au and $\langle e,i \rangle=0.027$.
The Mercury/Mars analogs more often form in these models, but they are typically too
massive, which affects the $S_{\rm c}$ and $S_{\rm d}$ statistics in the opposite way than in W11/{\tt case3}.

\subsubsection{Mars/Mercury}

$N_{\rm Mars}$ and $N_{\rm Merc}$ in Table 1 report the number of planets (per simulation) similar to Mars 
and Mercury. In each case, we report two numbers separated by a slash. The first number is the number of Mars/Mercury
analogs. The second number is the number of {\it strict} Mercury and {\it strict} Mars analogs as defined in Sect. 2.4. 
These numbers are reported only for the good simulations that produced exactly two Venus/Earth planets (Sect 2.4). 

Both Mercury and Mars obtained in our simulations are typically too massive and there is a significant variability among
random realizations of the same model. In the C98 model, the best results for Mars are obtained in the {\tt case1} and {\tt case3} 
models where the average Mars masses  are $M=0.15$ $M_{\rm Earth}$ and $M=0.09$ $M_{\rm Earth}$, respectively (Table 1). 
This can be compared to C98/{\tt nojov}, where the average Mars mass is $M=0.35$ $M_{\rm Earth}$ (and there are no strict Mars analogs). 
It shows that the giant planets could indeed help to reduce the mass of terrestrial planets that form at $r \sim 1.5$ au 
(Raymond et al. 2009, Clement et al. 2018). As for the strict Mars analogs, however, C98/{\tt jovend} performs at least as well as 
C98/{\tt case1} and C98/{\tt case3}. Specifically, in C98/{\tt jovend}, 50\% of `good' simulations produce strict Mars analogs 
(60\% produce good Earth/Venus planets, as indicated by $f_{\rm g}=0.6$ in the second column of Table 1, and of these 50\% have 
strict Mars analogs; thus, overall $0.6 \times 0.5 = 0.3$ of the C98/{\tt jovend} simulations match small Mars). This shows that the 
first-order effect on Mars mass comes from the existence of the giant planets (on their current orbits) and not from the 
instability itself, or the detailed behavior of the giant planets during the instability.

Comparing W11/{\tt jovend} with C98/{\tt case1}, as a proxy for a comparison of the narrow-annulus and instability models, we find 
that the results concerning the average mass of Mars and the number of strict Mars analogs show no resolvable differences 
(e.g., 0.4 strict Mars analogs for C98/{\tt case1} and 0.5 for W11/{\tt jovend}). It may therefore be difficult to favor one 
of these models over the other based on this criterion. As we pointed up above, however, the W11 model is clearly better in terms 
of the RMC and $\Delta a$ parameters than any C98 model ({\tt jovend}, {\tt case1}, etc.). This makes sense because when C98 
is combined with the instability model (or the jovian planets on their present orbits), small Mars can be born, but there is no 
mechanism in this model that would assure the tight radial spacing of Venus and Earth. In W11, where protoplanets start in the narrow 
annulus, the radial mass spreading is also a problem, but Venus and Earth tend to generally end up closer to each other. 

Mercury analogs obtained in our simulations are excessively massive. The best results for Mercury were seemingly obtained in 
the L15/{\tt case1} model (Fig. \ref{case1}) where the average mass of Mercury is $M=0.15$ $M_{\rm Earth}$ (Table 1), about three 
times larger than the actual mass. In this case, however, the results are based on only one simulation that produced good 
Venus/Earth analogs ($f_{\rm g}=0.1$ for L15/{\tt case1} in Table 1). In all other simulations, the average Mercury  mass is higher 
and the fraction of strict Mercury analogs is lower. The difficulty of our simulations to produce a really good Mercury analog 
with mass $\sim0.06$ $M_{\rm Earth}$ is pointing to some important limitation of our model, with the obvious suspect being our 
assumption of inelastic mergers during all collisions. 

\subsubsection{Asteroid belt}

Finally, we comment on the number of planetary-mass bodies ($M>0.01$ $M_{\rm Earth}$) stranded in the asteroid belt 
at the end of our simulations and on the overall depletion of main belt asteroids ($N_{\rm spl}$ and $f_{\rm ast}$; the last 
two columns in Table 1).\footnote{We define $f_{\rm ast}=N_{\rm ini}/N_{\rm end}$, where $N_{\rm ini}$ is the initial number of 
planetesimals with $2.2<a<3.5$ au and $N_{\rm end}$ is the final number of planetesimals in the asteroid belt region, as 
defined in Fig. \ref{real}.} In both cases, we report the average value per good simulation at $t=200$ Myr; this does not account 
for the removal of bodies after 200 Myr. The number of main belt asteroids is expected to be reduced by an additional factor 
of $\sim$2 over 4 Gyr (e.g., Minton \& Malhotra 2010, Nesvorn\'y et al. 2017). 
The models with $N_{\rm spl} \sim 1$ cannot therefore be easily ruled out because, with the additional depletion 
factor of $\sim$2, about a half of random model realizations would end up with no planets in the asteroid 
belt.\footnote{The stranded protoplanets could carve gaps in the asteroid belt which are not observed; this suggests 
$N_{\rm spl} \sim 0$.} The C98/{\tt nojov} model does not pass this filter by a large margin and is clearly 
inadequate. The W11/{\tt nojov} and L15/{\tt nojov} models with $N_{\rm spl} = 1.4$ and $N_{\rm spl} = 2.5$, respectively, 
are not ideal as well. The jovian planets help to reduce this problem, 
but even with them included, $N_{\rm spl}=1.5$ in the C98/{\tt jovend} and L15/{\tt jovend} models. More plausible 
results with $N_{\rm spl} \lesssim 1$ are obtained for all instability models, and for W11/{\tt jovend}. 

The asteroid belt depletion shows a strong correlation with the number of protoplanets at $r>2$ au. In the C98 model, 
where $\sim$50 protoplanets are placed on orbits between 2 and 4 au, the surviving fraction of main-belt asteroids varies 
between $f_{\rm ast}=0.007$ (C98/{\tt case2}) and 0.066 (C98/{\tt nojov}). The surviving fraction is much higher in the 
W11 and L15 models, where there are no protoplanets at $2<r<4$ au to start with. In these models, $f_{\rm ast}$ is lower when 
the giant planet instability is accounted for. Some 30-43\% of main-belt asteroids survive in the {\tt case1} and 
{\tt case2} models, which is comparable to the survival rate reported in Nesvorn\'y et al. (2017; again, one needs to factor 
in the additional depletion of $\sim$2 from the long-term dynamics). The overall depletion is larger in the {\tt case3} model, 
which was specifically designed to explain the orbital excitation of main belt asteroids (Deienno et al. 2018). Here we find 
$f_{\tt ast}=0.06$ for W11/{\tt case3} and $f_{\tt ast}=0.05$ for L15/{\tt case3} (Deienno et al. 2018 reported 
$f_{\tt ast}=0.09$ but did not have the terrestrial planets in their simulations).

\subsection{100-simulation sets for selected models} 

\begin{figure}
\epsscale{0.8}
\plotone{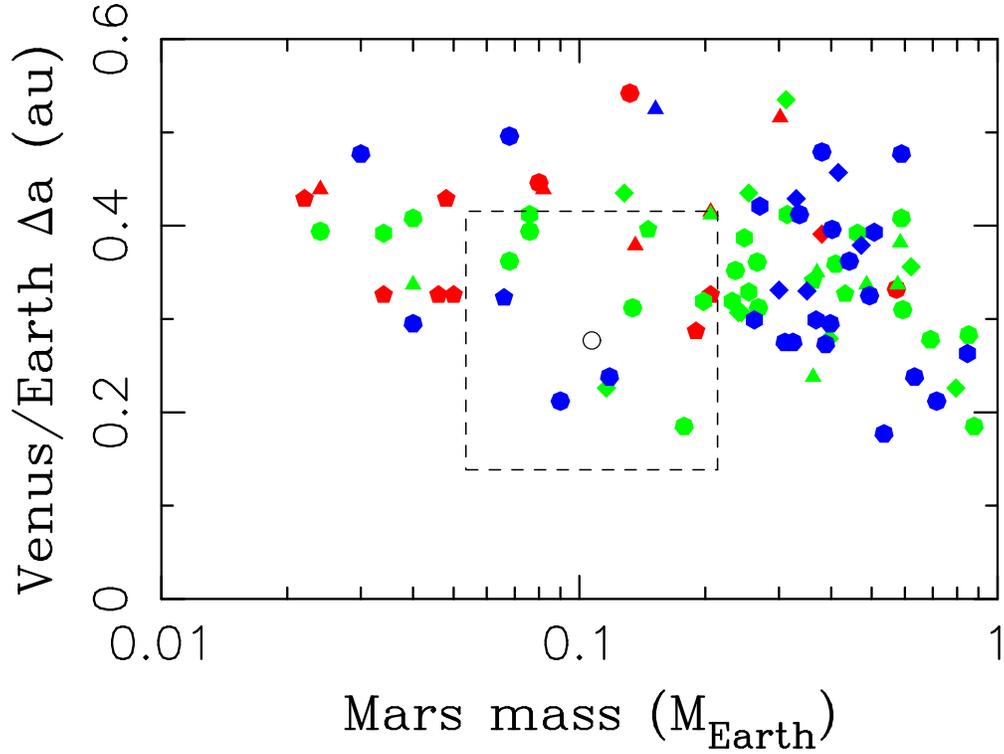}
\caption{The Mars mass ($M_{\rm Mars}$) and Venus/Earth separation ($\Delta a$) obtained in our 10-simulation sets. The symbol colors indicate 
the radial ranges of the initial protoplanet/planetesimal disks: red for C98, green for W11, and blue for L15. The shape of symbols defines
the migration/instability model: triangle for {\tt case1}, square for {\tt case2}, pentagon for {\tt case3}, hexagon for {\tt nojov}, heptagon 
for {\tt jovini}, and octagon for {\tt jovend}. The circle marks the real values: $M_{\rm Mars}=0.107$ $M_{\rm Earth}$ and $\Delta a = 0.277$ au. The 
dashed lines demarcate the region where $M_{\rm Mars}$ is within a factor of two to the real mass and the simulated $\Delta a$ is no more
than 1.5 times the real value and no less than 0.5 of the real value.}
\label{over}
\end{figure}

\begin{figure}
\epsscale{0.8}
\plotone{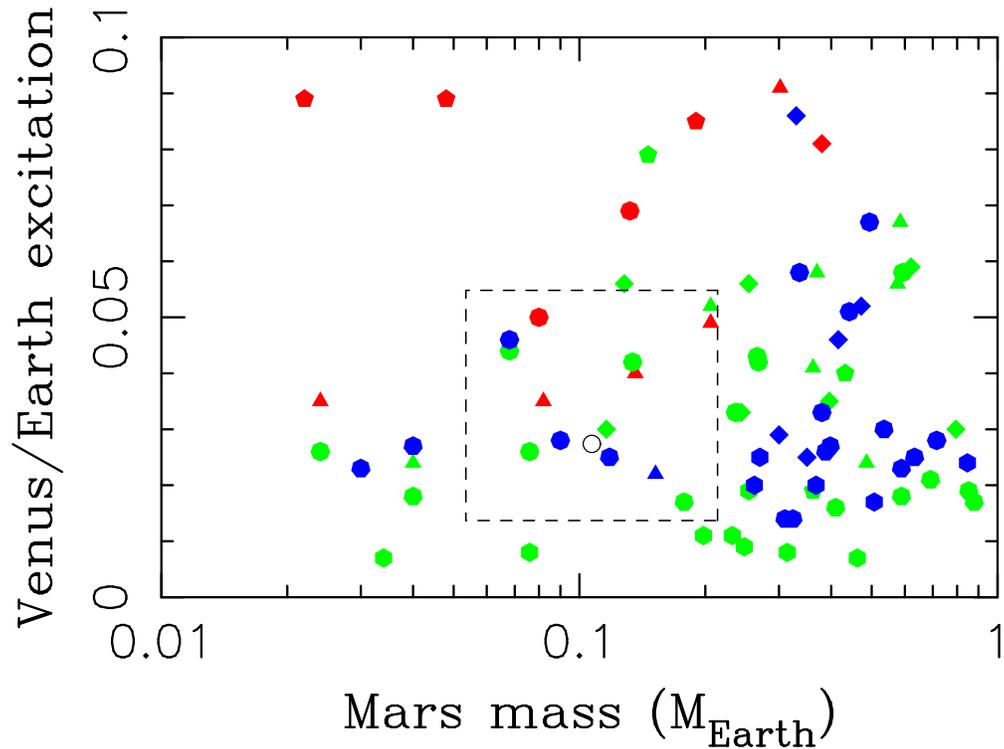}
\caption{The Mars mass ($M_{\rm Mars}$) and excitation of Venus/Earth orbits ($\langle e,i \rangle$) obtained in our 10-simulation sets. See
caption of Fig. \ref{over} for the description of different symbols. The circle marks the real values: $M_{\rm Mars}=0.107$ $M_{\rm Earth}$ and 
$\langle e,i \rangle=0.0274$ au. The dashed lines demarcate the region where $M_{\rm Mars}$ and $\langle e,i \rangle$ are within a factor 
of two to the real values.}
\label{over2}
\end{figure}

Figures \ref{over} and \ref{over2} summarize the results of simulations described in the previous text. 
We highlight the simulations that produced reasonable results: $M_{\rm Mars}=0.054$-0.214 $M_{\rm Earth}$ (within a factor of two to the real mass), 
$\Delta a=0.139$-0.415 au(no less than a half and no more than 1.5 times the real value) and 
$\langle e,i \rangle=0.0137$-0.0548 (within a factor of two to the real value). For C98, the best results were obtained for 
{\tt case1} where two of 10 simulations simultaneously satisfied these criteria (red triangles in the figures).
For W11, the best results were obtained for {\tt jovend} where three simulations satisfied the criteria (green octagons),
and for {\tt case1} and {\tt case2} (one successful simulation each). For L15, one simulation of {\tt jovend} looks 
good in Figs. \ref{over} and \ref{over2} (blue octagon), but that one ended with very small Earth/Venus analogs (just above
our $M>1/3$ $M_{\rm Earth}$ cutoff). 

Given the statistically small fraction of successful simulation, we select two promising models and perform 100 simulations
for each. The first model is a slight modification of W11/{\tt case1}, where planetesimals with $a>1.5$ au are placed
on orbits with larger eccentricities and larger inclinations. There are two reasons for this. First, the {\tt case1} 
instability, in absence of other effects, cannot explain the inclination excitation of main belt asteroids (Fig. \ref{s2orb}),
if $i \sim 0$ at $t_0$. Second, for W11 to be a closer proxy for the Grand Tack, the eccentricities and inclinations of 
planetesimals would have to be substantial at $t_0$. Specifically, we use the Rayleigh distributions with $\sigma_e=0.3$ and 
$\sigma_i=0.14$ for $a>1.5$ au (model W11e where `e' stands for `excited'). The second model is a modification of  
C98/{\tt case1}. We use a slightly steeper radial profile of protoplanets/planetesimals, $\gamma=1.5$, in an attempt to 
reduce the number of planets stranded in the asteroid belt (this model is called C98s in the following where `s' stands for
`steeper' profile). We also increase the total initial mass to 5 $M_{\rm Earth}$, splitting it equally between 
protoplanets and planetesimals, because the Venus/Earth analogs were somewhat too small in the original C98 setup 
(Fig. \ref{s1}). 

\begin{figure}
\epsscale{0.85}
\plotone{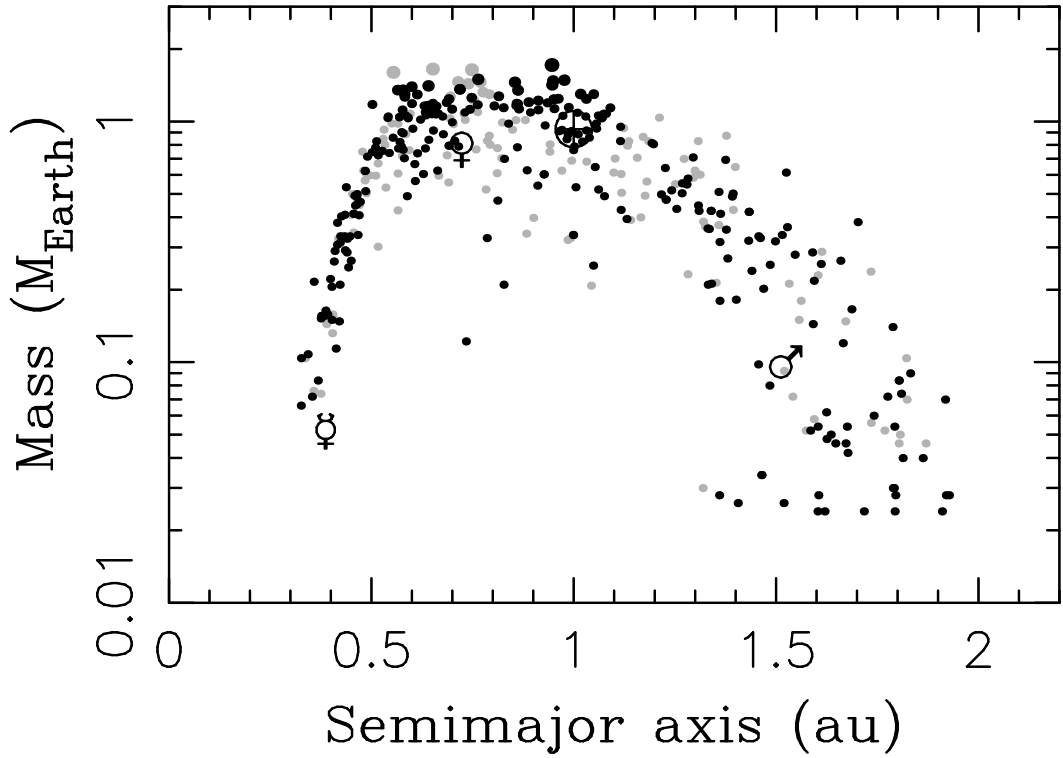}
\caption{The mass and radial distribution of protoplanets and planetesimals obtained in 100 simulations of the W11e/{\tt case1} model 
(Sect. 3.4). The black dots are the results with good Earth/Venus analogs (Sect. 2.4); all other simulations are denoted by gray dots. 
The real terrestrials planet are indicated by symbols: $\mercury$ for Mercury, $\venus$ for Venus, $\oplus$ for Earth, $\mars$ for Mars.}
\label{test4}
\end{figure}

\begin{figure}
\epsscale{0.65}
\plotone{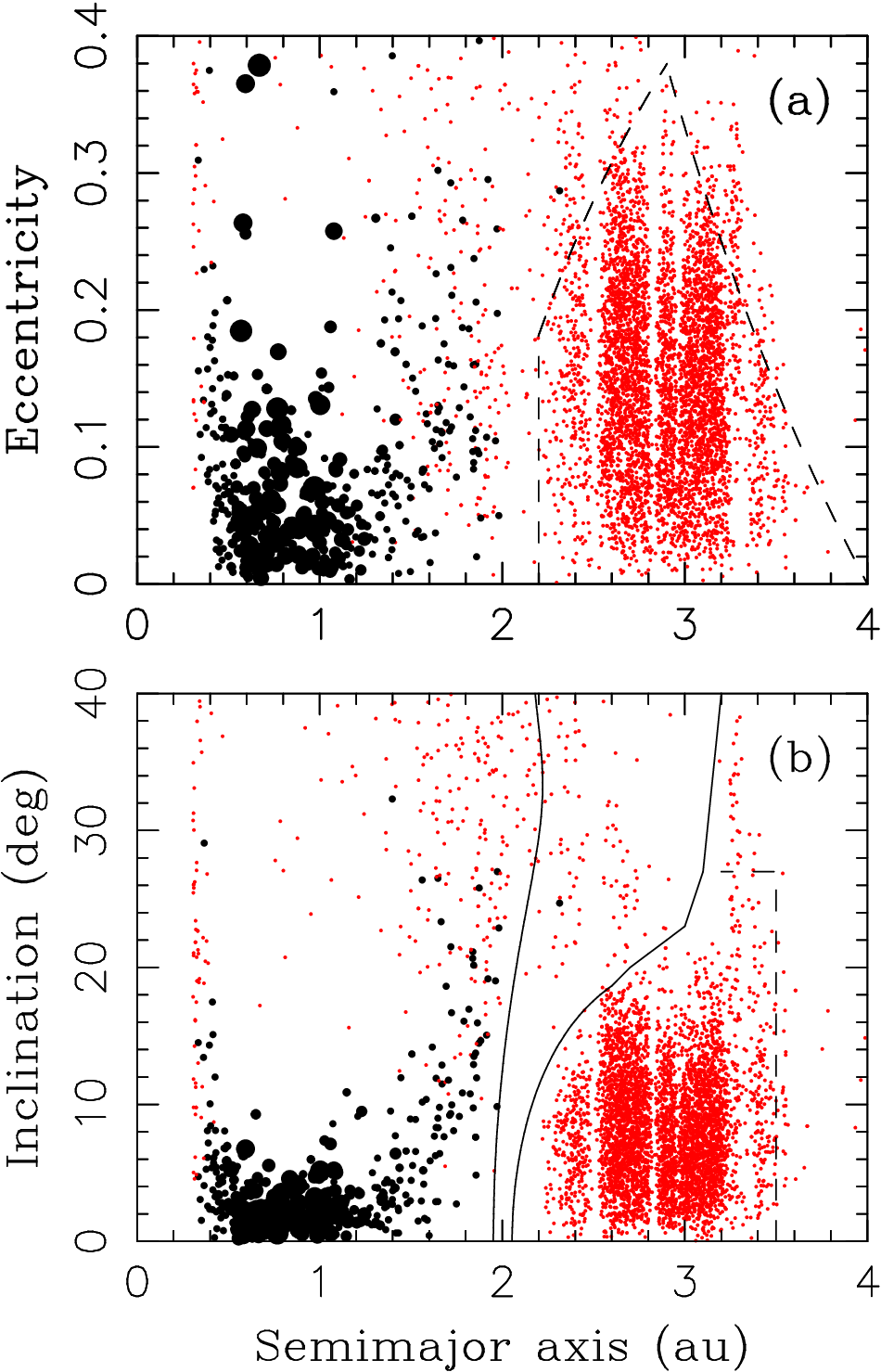}
\caption{The orbital distribution of planets and asteroids obtained in 100 simulations of the W11e/{\tt case1} model. 
The color highlights the mass of different bodies: planets with $M>0.01$ $M_{\rm Earth}$ are shown by black symbols,
leftover planetesimals and asteroids with $M<0.01$ $M_{\rm Earth}$ are shown by red symbols.}
\label{test4o}
\end{figure}

\begin{figure}
\epsscale{0.65}
\plotone{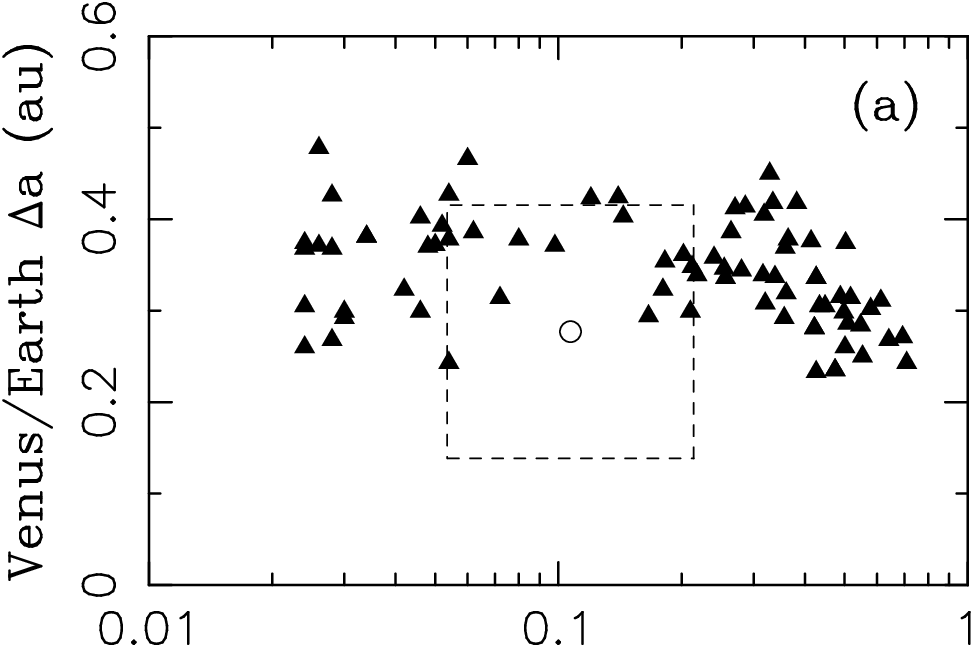}\\[2.mm]
\plotone{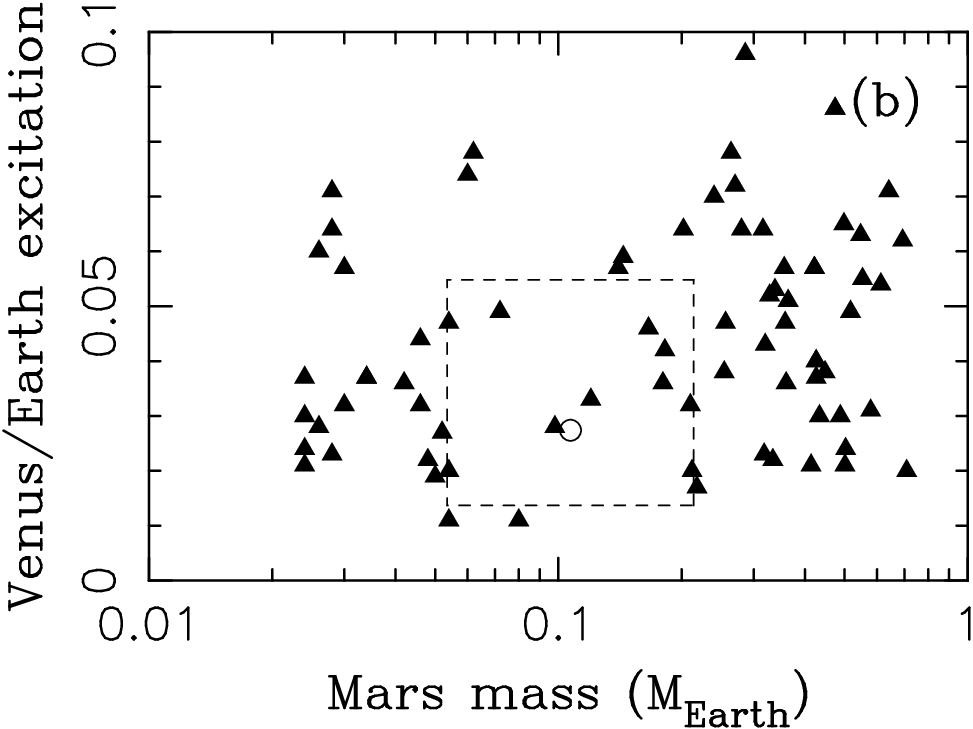}
\caption{(a) The Venus/Earth separation ($\Delta a$) and (b) excitation of Venus/Earth orbits ($\langle e,i \rangle$) obtained in 100 
simulations of W11e/{\tt case1} (only good systems plotted here). See captions of Figs. \ref{over} and \ref{over2} for additional 
information.}
\label{over_test4}
\end{figure}

The results for the W11e/{\tt case1} model are documented in Figs. \ref{test4}--\ref{over_test4} and Table 1. Overall, 
the results are similar to those obtained (with smaller statistics) for the W11/{\tt case1} model. The radial
mass distribution of final planets has the desired profile (Figs. \ref{test4}), but the Mercury analogs are again somewhat 
too massive. The orbital excitation of Venus/Earth planets in W11e/{\tt case1} is slightly larger than in W11/{\tt case1} 
(Table 1), probably because the planetesimals beyond 1.5 au had excited orbits in W11e, and could not damp the planetary 
eccentricities and inclinations as efficiently as in W11 (via a secular coupling). The average Mars mass 
in W11e/{\tt case1} is $0.26$ $M_{\rm Earth}$, about 2.4 times higher than what would be ideal, but this is slightly biased 
by the fact that planets at 1.2-1.5 au tend to be more massive (Fig. \ref{test4}). If a more strict selection of Mars 
analogs were made, the average mass would be lower (e.g., $M \simeq 0.2$ $M_{\rm Earth}$ for $1.4<a<1.6$ au). There are 0.21 
of strict Mars analogs per good simulation, which is an improvement from W11/{\tt case1}, where we did not have the 
needed resolution to identify any strict Mars analogs.

Overall, 8\% of W11e/{\tt case1} simulation 
produced satisfactory results for the Mars mass, $\Delta a$ and $\langle e,i \rangle$ (as defined by dashed rectangles in 
Figs. \ref{over_test4}). This may seem small, but if the probability of satisfying each criterion were 50\%, about 13\% 
cases would be expected to satisfy them all (assuming no correlations). From Fig. \ref{over_test4}a we see that the most
persistent problem remains to match $\Delta a$. All but one satisfactory simulations (within the dashed rectangles in
Figs. \ref{over_test4}) show $\Delta a$ values that are higher than $\Delta a=0.277$ au of the real planets.

The orbital structure of the asteroid belt in W11e/{\tt case1} looks great (Fig. \ref{test4o}). About 11\% of main belt 
asteroids survive by $t=200$ Myr, corresponding to a factor of $\sim$3 larger dynamical depletion than in the W11/{\tt case1} 
(due to the initially excited orbits of asteroids in W11e).\footnote{Here the depletion factor is referred to the population 
of planetesimals starting with $2.2<a<3.5$ au and ending in the main belt. Note that only a fraction of planetesimals with 
$2.2<a<3.5$ au is actually in the main belt region at $t_0$ (as defined in Fig. \ref{real}), because eccentricities and 
inclinations are relatively high in W11e.} Assuming that the asteroid belt was depleted by an additional factor of $\sim 2$ in 
the 0.2-4.5 Gyr interval (e.g., Minton \& Malhotra 2010, Nesvorn\'y et al. 2017), we find $f_{\rm ast} \sim 0.055$ overall.  
For comparison, Deienno et al. (2016) reported that $\sim$5\% of bodies starting with $1.8<a<3.6$ au survive in their 
simulations over 4.5 Gyr.

The current mass of the asteroid belt is $\simeq 5 \times 10^{-4}$ $M_{\rm Earth}$. Assuming that the total depletion factor over
4.5 Gyr is $f_{\rm ast}=0.05$, we estimate that the initial mass at 2.2-3.5 au and $t_0$ was $5 \times 10^{-4} / 0.05=
10^{-2}$ $M_{\rm Earth}$. This is a factor of $\sim 100$ lower than the mass inferred from the Minimum Mass Solar Nebula (MMSN;
Weidenschilling 1977, Hayashi 1981). This means that: (1) $\sim 100$ of the population was depleted prior to $t_0$, 
which would be consistent with Jupiter's inward-then-outward migration in the Grand Tack model (Walsh et al. 2011), (2) 
the planetesimal formation at 2.2-3.5 au was relatively inefficient (e.g., Raymond \& Izidoro 2017), perhaps due to the 
early formation of Jupiter (Kruijer et al. 2017), or (3) the MMSN is not a good reference (e.g., observed protoplanetary 
disks often show radial gaps; Andrews 2020). 

Finally, we briefly comment on C98s/{\tt case1}. This model is not successful for several different 
reasons. The planetary mass distribution is similar to that in W11e/{\tt case1} but the Mercury analogs are far too massive due 
to the steeper radial profile of protoplanets (initially extending down to $r_{\rm in}=0.3$ au). About one planetary-mass body 
is stranded in the asteroid belt at $t=200$ Myr in C98s/{\tt case1}, which is worse than what we found in C98/{\tt case1}: 
the slightly steeper radial profile did not help. The orbital excitation of Venus/Earth is too large. In fact, all 41 good 
simulations in C98s/{\tt case1}   
ended with $\langle e,i \rangle > 0.03$. The average AMD is $S_{\rm d}=0.0068$, about 3.8 times larger that the real value. As for
the RMC and radial separation of Venus and Earth, we find average $S_{\rm c}=24.4$ and average $\Delta a=0.375$ au, again
significantly larger than the real values. Interestingly, unlike in Fig. \ref{over_test4}a where the $\Delta a$ values are 
concentrated near 0.3-0.4 au in W11e/{\tt case1}, here $\Delta a$ ranges from 0.2 to 0.6 au, and a small fraction of cases
yield the right orbital separation. The smaller variability of $\Delta a$ (and $S_{\rm c}$) in W11e/{\tt case1} is most likely 
related to the initial distribution of protoplanets in the narrow annulus.    

\subsection{Simulations starting with Mars-size protoplanets at $t_0$}

Jacobson \& Morbidelli (2014) suggested that the RMC can be enhanced in the narrow annulus model, or equivalently $\Delta a$ 
can be reduced, if the late stage of the terrestrial planet accretion started with Mars-size protoplanets at $t_0$. 
If the initial protoplanets are more massive, they can grow faster by accretional collisions, and there is presumably not 
enough time for them to radially spread. To test this effect, we perform additional 100 simulations for both
the W11e/{\tt case1} and C98s/{\tt case1} models, but this time starting with Mars-mass protoplanets at $t_0$ (20 protoplanets 
in W11e, 25 protoplanets in C98; these jobs are called W11e/20M and C98s/25M; Table 1). We find that the C98s/25M
model does not offer any benefits over the C98 or C98s models. The radial mass distribution of the final planets in 
C98s/25M is relatively flat and there are no strict Mars analogs (Table 1). For brevity, we do not discuss C98s/25M here.
   
We find that the RMC values are generally higher, in agreement with Jacobson \& Morbidelli (2014), and the AMD values are 
generally lower in the W11e/20M model (Table 1). Specifically, we obtain average $S_{\rm c}=51.2$ in W11e/20M, to be compared 
with $S_{\rm c}=46.4$ in the standard W11e/{\tt case1} and real $S_{\rm c}=89.9$, and $S_{\rm d}=0.0034$ in W11e/20M, to be compared 
with $S_{\rm d}=0.0040$ in the standard W11e/{\tt case1} and real $S_{\rm d}=0.0018$. The differences between the model and real 
values are still large but this is influenced by Mercury/Mars. As we discussed in Sect. 2.4, if Mercury/Mars form too 
massive in the simulations, this affects $S_{\rm c}$ and $S_{\rm d}$. This problem is exacerbated in W11e/20M where 
the protoplanets are massive at $t_0$ and Mercury/Mars end up to be more massive on average (Table 1).\footnote{Jacobson \& 
Morbidelli (2014) reported $S_{\rm c}$ values that are a better match to the real $S_{\rm c}=89.9$. This is most like due to 
their initial setup where protoplanets and planetesimals start with $r>0.7$ au. With this setup, Mercury analogs with $a<0.5$ 
au do not typically form in their simulations and therefore do not influence the $S_{\rm c}$ statistics. In our W11e models, 
the original planetesimal disk extends down to $r=0.3$ au, massive Mercury analogs with $a<0.5$ au form (Fig. \ref{test4}), 
and this reduces the RMC.} It may thus seem better to utilize $\Delta a$ and $\langle e,i \rangle$ instead.

\begin{figure}
\epsscale{0.8}
\plotone{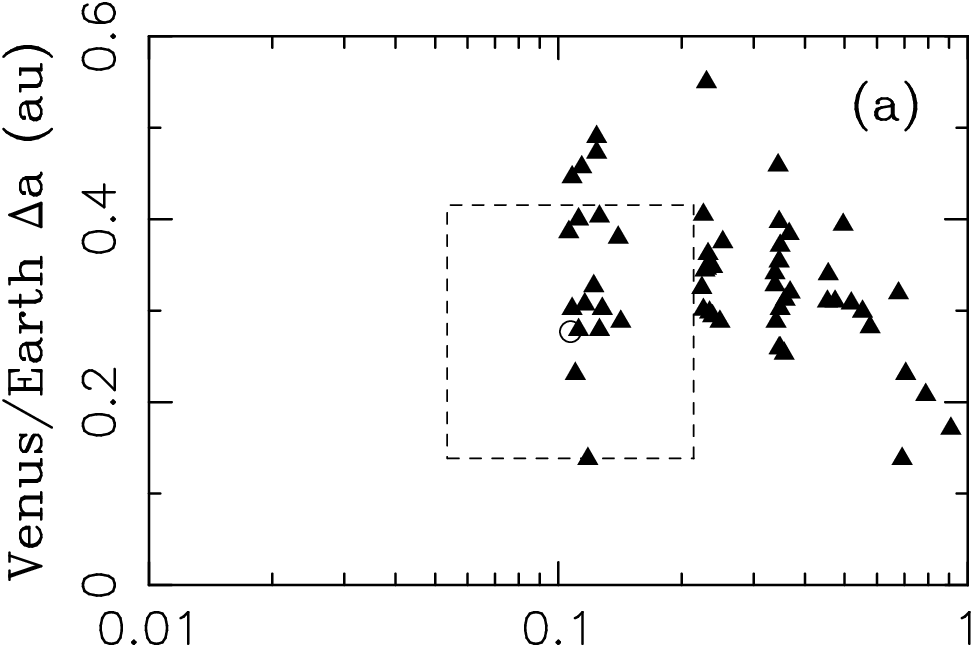}\\[2.mm]
\plotone{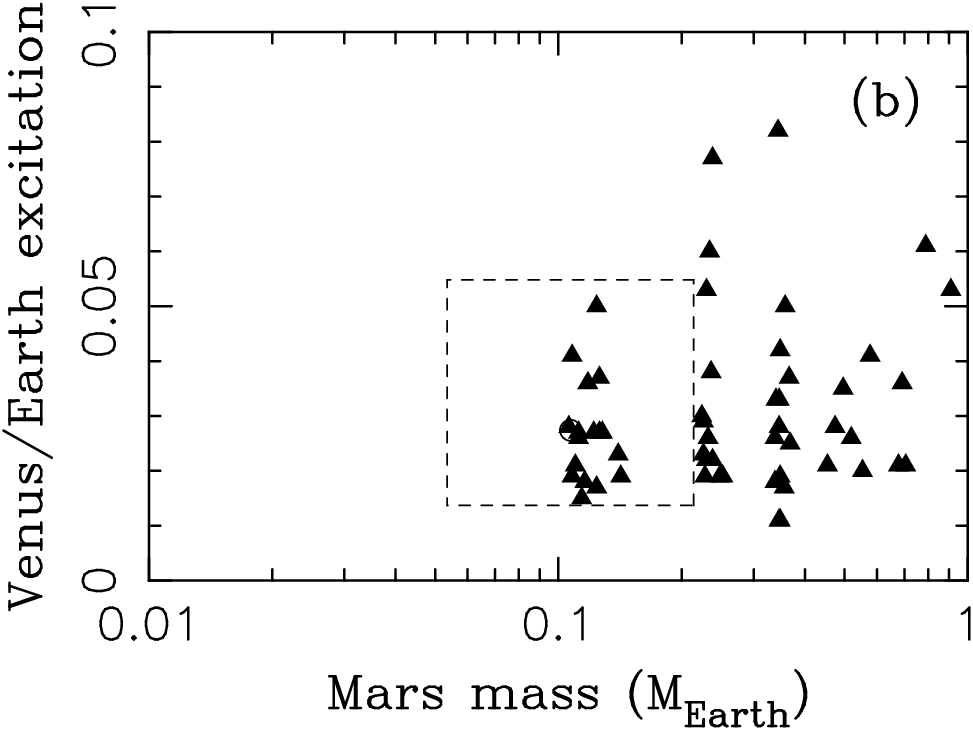}
\caption{(a) The Venus/Earth separation ($\Delta a$) and (b) excitation of Venus/Earth orbits ($\langle e,i \rangle$) obtained in 100 
simulations of W11e/20M (only good systems plotted here). See captions of Figs. \ref{over} and \ref{over2} for additional information.}
\label{mars1}
\end{figure}

The average $\Delta a$ values are practically the same in the W11e and W11e/20M models ($\Delta a \simeq 0.34$ au, Table 1).
We therefore do not see much influence of the starting mass of protoplanets on the final separation of Venus/Earth. Compared 
to the real $\Delta a \simeq 0.28$ au, the average model values are only $\sim$20\% higher (whereas the average model 
$S_{\rm c}$ values discussed above are 40\% lower that the real $S_{\rm c}=89.9$). The range of $\Delta a$ values is larger
in W11e/20M ($\simeq0.15$-0.5; Fig. \ref{mars1}a) than in W11e ($\simeq0.25$-0.4; Fig. \ref{over_test4}a) probably because 
the accretion of Venus/Earth is more stochastic if protoplanets start more massive. If protoplanets start with roughly
the Mars mass, in nearly 20\% of simulations of W11e/20M, the strict Mars analogs are scattered from $r<1$ au to 1.2-1.8 au 
and only grow by accreting planetesimals. They end up having masses 1-1.5 $M_{\rm Mars}$ (Fig. \ref{mars1}). Unlike
in Fig. \ref{over_test4}a, where the W11e model did not no produce any really close matches to the real terrestrial 
planets (model Mars was either too small or too big, or model $\Delta a$ was too large), there are some very good matches
in Fig. \ref{mars1}a. 

The initial mass of protoplanets affects $\langle e,i \rangle$. We obtain $\langle e,i \rangle=0.053$ in the W11e model 
and $\langle e,i \rangle=0.039$ in the W11e/20M model (to be compared with real $\langle e,i \rangle=0.027$). This is a 
significant improvement. Moreover, if we only count the simulations where a Mars analog formed in W11e/20M (56\% of cases; 
Fig. \ref{mars1}b), we find average $\langle e,i \rangle=0.032$ -- only 15\% higher than the real value. In addition, if 
we only count the cases with strict Mars analogs ($M<0.2$ $M_{\rm Earth}$; 17\% of cases; Fig. \ref{mars1}b), then 
$\langle e,i \rangle=0.027$, which is identical to the real value. This trend is not as strong in the W11e where 
counting the simulations with a Mars analog gives $\langle e,i \rangle=0.046$. We therefore endorse the suggestion of 
Jacobson \& Morbidelli (2014) that it is
beneficial to initiate the simulations with Mars-mass protoplanets.\footnote{It has to be noted that adopting
Mars-mass protoplanets at $t_0$ limits the predictive power of the late-stage accretion model, because Mars does 
not grow in simulations; it is assumed to be practically fully grown at $t_0$. This also raises questions about whether 
even more massive planets could have formed at $\sim 1$ au before the gas disk dispersal (e.g., Johansen et al. 2015, 
Bro\v{z} et al. 2020).}     

\begin{figure}
\epsscale{0.8}
\plotone{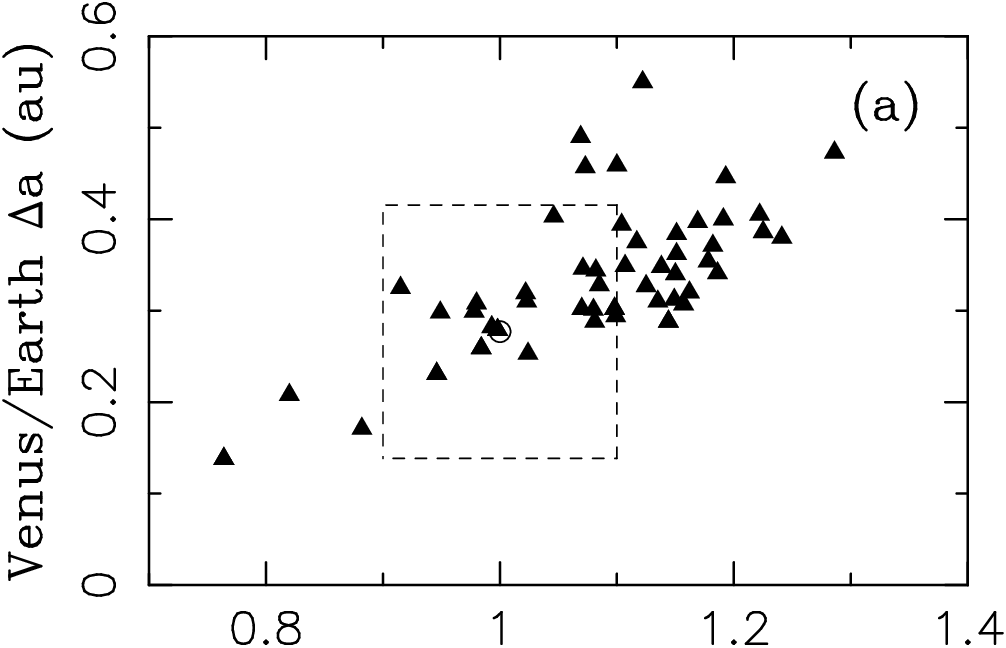}\\[2.mm]
\plotone{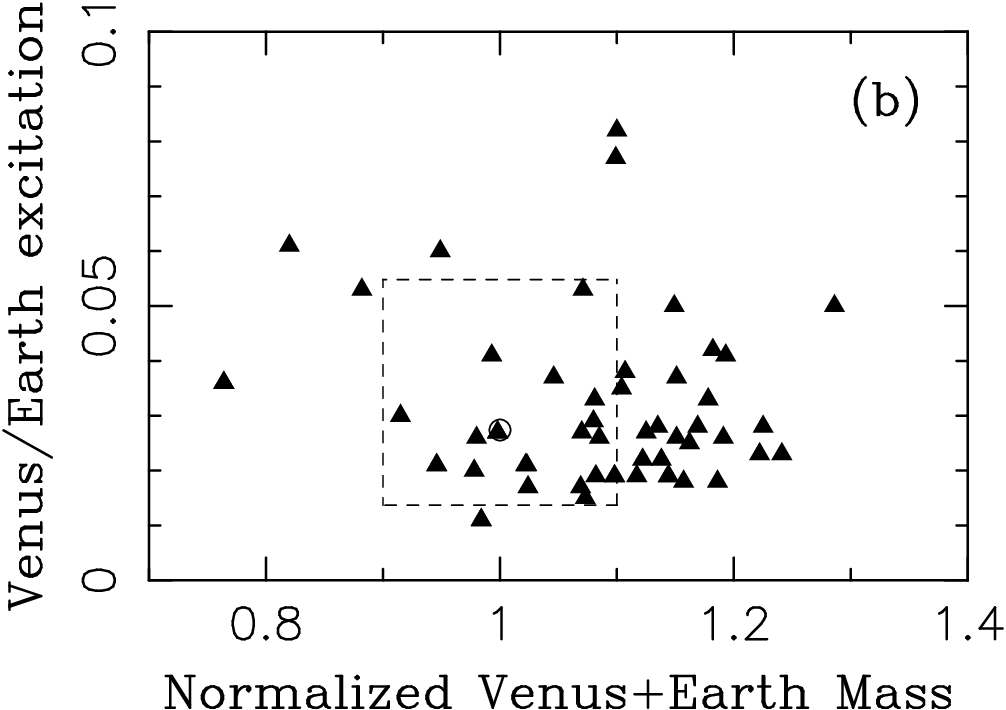}
\caption{(a) The Venus/Earth separation ($\Delta a$) and (b) excitation of Venus/Earth orbits ($\langle e,i \rangle$) obtained in 100 
simulations of W11e/20M, here plotted as a function of the normalized total mass of Venus/Earth planets ($M_{\rm VE}$).
See captions of Figs. \ref{over} and \ref{over2} for additional information.}
\label{mve}
\end{figure}

Finally, we test possible trends with the mass of Venus/Earth analogs. For that, we compute the total mass of the Venus/Earth 
planets in the good systems and normalize it by the total mass of real Venus and Earth, $M_{\rm VE}=(M_1+M_2)/M_{\rm real}$ with 
$M_{\rm real}=1.82$ $M_{\rm Earth}$. We find that both the W11e and W11e/20M models give average $M_{\rm VE}=1.07$ (i.e., the mass 
is only 7\% higher than the actual value). This is typically caused by a slightly larger mass of the Venus analogs (Fig. 
\ref{test4}). The radial separation of Venus/Earth planets in W11e/20M correlates with their total mass (Fig. \ref{mve}a) 
but the orbital excitation does not (Fig. \ref{mve}b). If we restrict the range to strict Venus/Earth analogs, $0.9<M_{\rm VE}<1.1$, 
and only count systems where a Mars analog formed, we obtain $\Delta a = 0.317$ au and $\langle e,i \rangle=0.030$ in 
W11e/20M, which is a significant improvement over the values listed in Table 1. We thus see that obtaining the correct planetary 
masses in the simulations links to better results for $\Delta a$ au and $\langle e,i \rangle$ as well. Such a trend is not 
expected a priori. The trends for W11e are similar but the overall results are not as good (e.g., $\langle e,i \rangle=0.044$ 
for $0.9<M_{\rm VE}<1.1$ with a Mars analog).  


\subsection{The best case}

\begin{figure}
\epsscale{0.75}
\plotone{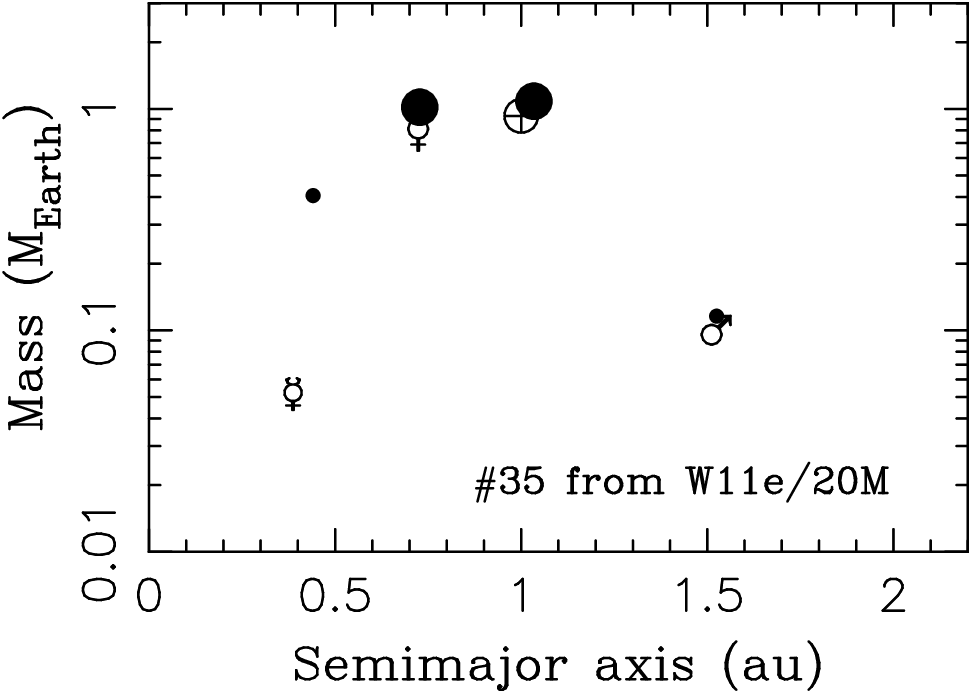}
\caption{The mass and radial distribution of planets in our best simulation of W11/20M with the {\tt case1} instability.
The size of a dot increases with planet mass. The real terrestrial planets are indicated by planetary symbols.}
\label{job35}
\end{figure}

\begin{figure}
\epsscale{0.65}
\plotone{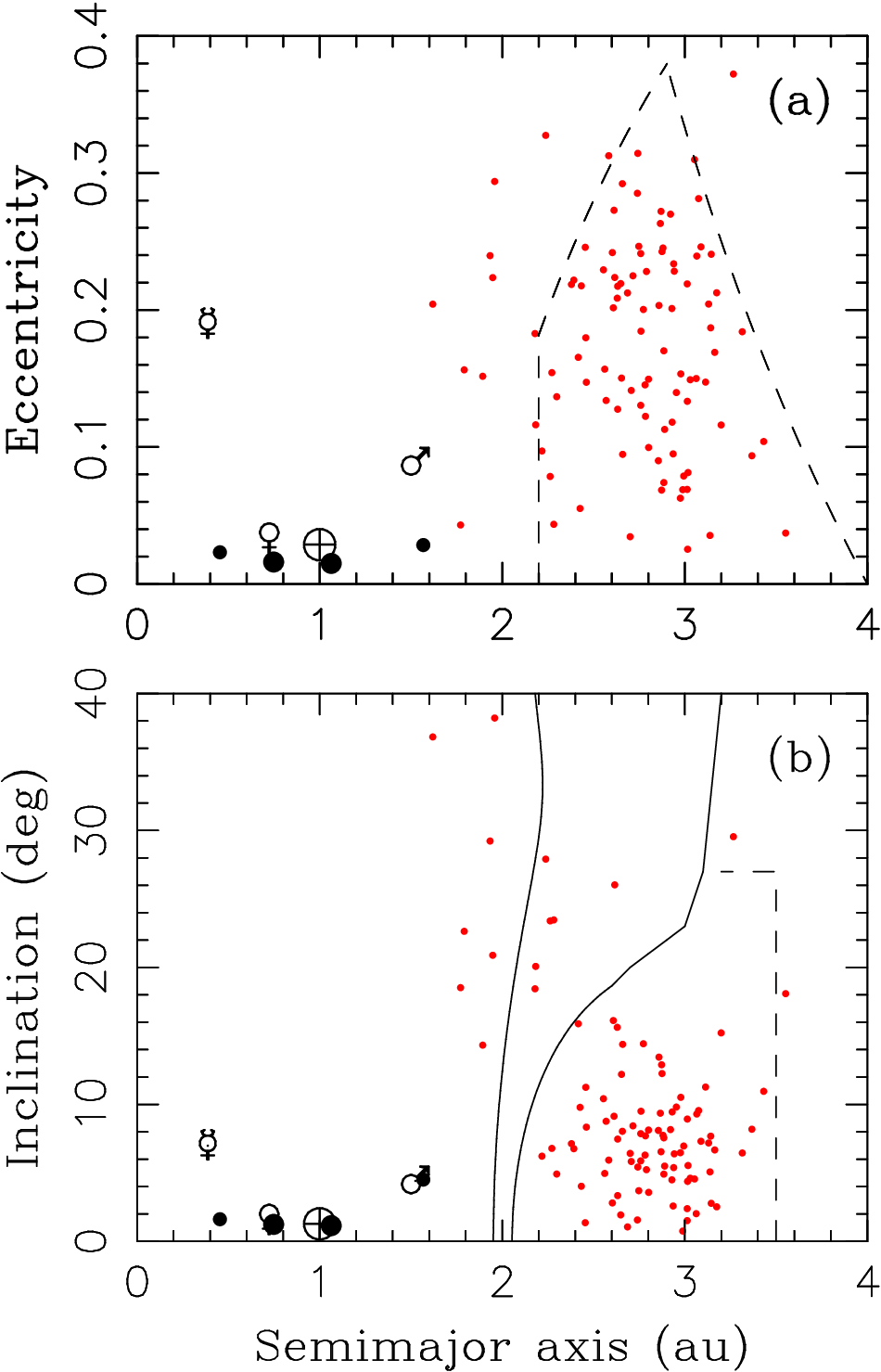}
\caption{The orbits of planets (black dots) and asteroids (red dots) obtained in our best simulation of W11/20M with the {\tt case1}
instability. The real terrestrial planets are indicated by planetary symbols. See caption of Fig. \ref{real} for the meaning of lines. 
The mean orbital elements of planets were computed over the last $10^8$ yr.}
\label{job35_orb}
\end{figure}

Here we highlight a case from the W11e/20M model (job number 35) that produced a particularly good match to the terrestrial
planets (Figs. \ref{job35} and \ref{job35_orb}). Venus, Earth and Mars formed in this simulation with nearly perfect masses 
and semimajor axis: 
model 1.017 $M_{\rm Earth}$ and $a=0.727$ au for Venus (0.815 $M_{\rm Earth}$ and $a=0.723$ au),
model 1.083 $M_{\rm Earth}$ and $a=1.034$ au for Earth (1.0 $M_{\rm Earth}$ and $a=1.0$ au),
model 0.116 $M_{\rm Earth}$ and $a=1.525$ au for Mars (0.107 $M_{\rm Earth}$ and $a=1.524$ au), where the numbers in parentheses
list the real values for reference. The separation of Venus and Earth is excellent ($\Delta a = 0.0307$ au vs. real 
$\Delta a = 0.0277$ au). Here the main difference is that Venus ended with the mass about 20\% higher than what would be ideal. 
The Mercury analog is very massive but that may be related to our assumption of inelastic mergers, as we discussed previously. 
We therefore do not comment on Mercury below.

The model orbits of Venus and Earth are slightly less excited than in the real system, resulting in $\langle e,i \rangle=0.0176$ (to 
be compared with real $\langle e,i \rangle=0.0274$). This is not bad. In fact, it is encouraging to see, for a change, a case 
with the low AMD. The model orbit of Mars has a nearly perfect inclination (mean $i=4.51^\circ$ vs. real mean $i=4.41^\circ$) 
and somewhat lower eccentricity (mean $e=0.028$ vs. real mean $e=0.069$). We would probably need to perform a larger suite of 
simulations, possibly with a less massive planetesimal population (Jacobson \& Morbidelli 2014), to obtain a better match to Mars's 
eccentricity. The orbits of main belt asteroids look good (compare Figs. \ref{job35_orb} and \ref{real}). The model orbital 
inclinations are slightly lower but this is related to our choice of the initial conditions and could easily be remedied by 
adopting a slightly wider initial distribution.

\begin{figure}
\epsscale{0.75}
\plotone{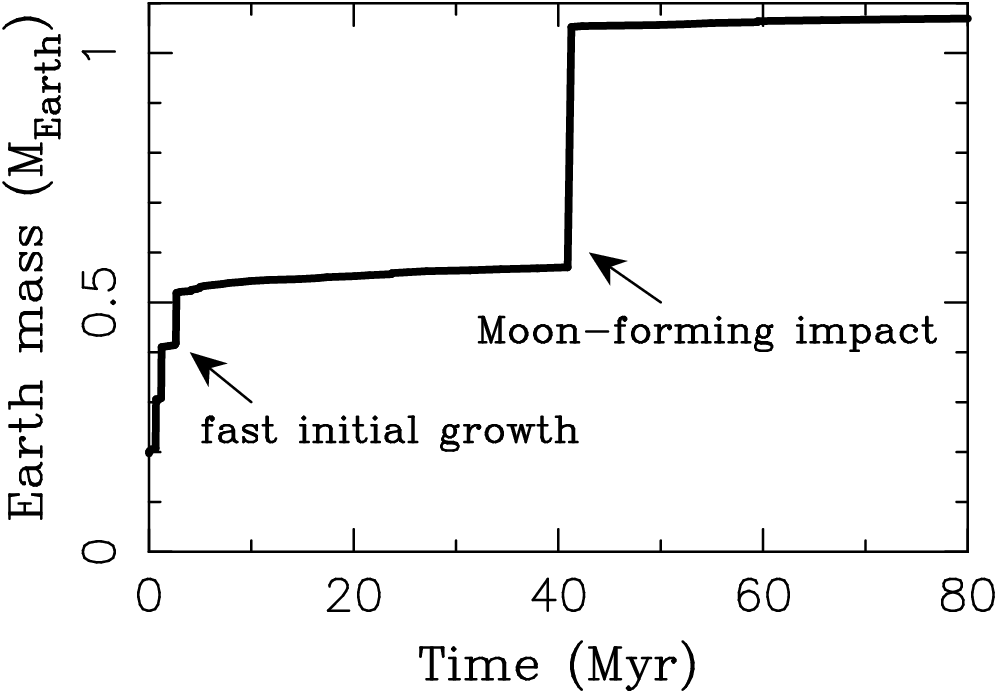}
\caption{The accretional growth of Earth in our best simulation of the W11/20M/{\tt case1} model. The Moon-forming collision
between two roughly equal-mass bodies occurred at $t=41.3$ Myr.}
\label{growth}
\end{figure}

A spectacular feature of the job highlighted here is related to the Earth's accretion history and the Moon-forming impact (Fig. 
\ref{growth}). The initial growth of proto-Earth is fast. After accreting several Mars-class protoplanets, the proto-Earth reaches 
mass $\simeq0.52$ $M_{\rm Earth}$ at $t \simeq 3$ Myr after the start of our simulation. A prolonged stage of planetesimal accretion
follows during which the proto-Earth modestly grows to $\simeq0.57$ $M_{\rm Earth}$ by $t \simeq 40$ Myr. Shortly after that, at 
$t=41.3$ Myr, an accretional collision between two roughly equal-mass bodies occurs, and the Earth mass shoots up to 
$\simeq 1.05$ $M_{\rm Earth}$.\footnote{The last planetesimal impact on the Earth occurred at $t \simeq 84$ Myr in our simulation but this 
reflects the resolution used here rather than the actual termination of the terrestrial bombardment.} For a comparison, Venus grows 
to $\simeq$85\% of its final mass in only 3 Myr after the start of the simulation; the remaining $\simeq$15\% of mass is supplied 
to Venus by planetesimals over an extended time interval ($\sim100$ Myr). This gives support to the suggestion of Jacobson et al. (2017)
who argued that Venus has not developed a persistent magnetic field because it did not experience any late energetic impacts that 
would mechanically stir the core and create a long-lasting dynamo.

The accretion growth of the Earth shown in Fig. \ref{growth} would satisfy constraints from the Hf-W and U-Pb isotope systematics,
assuming that the degree of metal-silicate equilibration was intermediate between the full and 40\% equilibration considered 
in Kleine \& Walker (2017; their figure 8). The last major accretion event at $t=41.3$ Myr corresponds to a very low 
speed collision (the impact-to-escape speed ratio $v_{\rm imp}/v_{\rm esc}=1.01$) between two nearly equal-mass protoplanets 
(the impactor-to-total mass ratio $\Gamma=0.46$). This falls into the preferred regime of Moon-forming impacts investigated in 
Canup (2012). In particular, the collision of two protoplanets with $\Gamma>0.4$ would lead to the Moon formation from a well 
mixed disk and would therefore imply matching oxygen-isotope compositions of the Earth and the Moon. The two protoplanets 
involved in the Moon-forming collision were nearly fully formed and on their pre-impact orbits ($a=0.95$ and 1.11 au; radial 
separation $\simeq0.16$ au) already by $t \simeq 3$ Myr after $t_0$. Their orbits intermittently crossed each other during 
orbital eccentricity oscillations, but it took nearly 40 Myr before the two protoplanets actually collided. A similar case was 
investigated in DeSouza et al. (2020). This demonstrates that the Moon-forming collision could have happened tens of Myr after 
the giant planet instability. 

\section{Conclusions}

The main conclusions of this work are:

\begin{enumerate}
\item We find that the early giant planet instability excites and depletes the population of protoplanets at $r \sim 1.5$ au and 
stalls Mars growth (Clement et al. 2018). The standard dynamical instability models preferred from previous work (Nesvorn\'y 2018) 
work just fine for that (stronger instabilities are not required). In fact, even if the giant planets are placed on their present 
orbits at $t_0$ and included in simulations of the terrestrial planet accretion, small Mars can form. The chief effect on 
Mars's growth is thus the presence of the giant planets and not the instability itself or the detailed orbital behavior of planets 
during the instability.
\item Some of the results reported in this work may suggest that the Mars mass and other properties of the terrestrial planet system
depend on the behavior of the giant planets. For example, when the giant planets are placed on their current orbits at $t_0$, 
the transition from the high to low planet mass at $\sim 1.5$ au is steeper than if the giant planets realistically evolve 
in one of our instability cases (Fig. \ref{s1}). Differences between different instability cases (other than the dependence on 
$t_{\rm inst}$; see below) may exist as well (Table 1), but we are unable to resolve them in detail with only 10 simulations done 
for each base model. Ideally, we should perform additional simulations of each model and vary $t_{\rm inst}$ (Clement et al. 
2018). This would help to distinguish between the effect of $t_{\rm inst}$ and the detailed behavior of the giant planets 
during the instability. This effort is left for future work.
\item Simulations of C98/{\tt case2} (the classic extended disk model of Chambers \& Wetherill (1998) with 
the {\tt case2} instability; Fig. \ref{fig2}) and L15/{\tt case2} (a model with the limited protoplanet growth for $r \gtrsim 
1.5$ au; Levison et al 2015) with $t_{\rm inst}=20$ Myr produced Mars planets that are too 
massive (average $M \simeq 0.37$ $M_{\rm Earth}$) and no strict Mars analogs with $M < 0.2$ $M_{\rm Earth}$. This shows that, if the 
giant planets are indeed responsible for small Mars, they would have to reach their current orbits early, some $t<10$ Myr 
after the gas disk dispersal (as suggested by Clement et al. 2018). Mars would become too massive if $t_{\rm inst}>10$~Myr. 

\item When the narrow annulus model (Hansen 2009) is combined with the early giant planet instability, the mass of Mars is 
slightly reduced, but the overall results do not change much (Clement et al. 2018 reached the same conclusion). Thus, if the 
terrestrial protoplanets indeed formed from the narrow annulus, the small mass of Mars is primarily the consequence of that, and 
not of the giant planet instability. If the terrestrial protoplanets did not form from the narrow annulus, the giant planet 
would help to create small Mars, assuming they reached their current orbits early (see above), but this model shows 
deficiencies elsewhere (e.g., weak radial mass concentration).
\item The orbital coupling during the giant planet instability could stall Mercury's growth and excite its orbital eccentricity and 
inclination. Roig et al. (2016) already demonstrated the orbital excitation of Mercury for the late instability model. Still, 
the Mercury mass obtained in our model with inelastic mergers tends to be too high, probably indicating that a more realistic
treatment of collisions is needed (e.g., Clement et al. 2019b). There is a trend with more massive Mercury analogs forming 
too close to Venus and less massive Mercury analogs forming at approximately the right location (e.g., Fig. 14). This probably 
reflects different histories of the scattering encounters between proto-Venus and proto-Mercury in individual simulations. 
\item The asteroid belt structure provides and interesting constraint on the terrestrial planet formation and the giant planet instability.
The models with shallow radial profiles of protoplanets, and no truncation, tend to leave $\sim 1$ planetary-size body stranded in the 
asteroid belt at $t=200$ Myr after $t_0$. The orbital excitation of main belt asteroids can date back to times before $t_0$ (Walsh et al. 
2011, Deienno et al. 2016) or after $t_0$ (Deienno et al. 2018, Clement et al. 2019a).
\item In one of our best models (W11e/{\tt case1}), $\simeq$8\% of simulations simultaneously reproduce small Mars (within a factor of 2 in 
mass), the tight orbital spacing of Venus and Earth (no more than 1.5 times the actual value), and correct orbital excitation (within a factor of 2).
It is thus possible that a correct combination of processes have already been identified but the real terrestrial planet system slightly deviates 
from a typical simulation outcome (slightly larger Mars, lower RMC, etc.). It is still intriguing, however, that 17 out of our 100 simulations 
of W11e/{\tt case1} (the narrow annulus model with the {\tt case1} instability, Fig. \ref{fig1}; Walsh et al. 2011)
produced roughly the right Mars mass (within a factor of two), but of these only one ended with $\Delta a$ 
smaller than the real $\Delta a=0.277$ au (Fig. \ref{over_test4}). This may signal a systematic problem with our understanding of the 
terrestrial planet formation.  
\item The radial mass concentration of the terrestrial planets, and specifically the tight radial separation of Venus and Earth, are
the most troublesome features of match. The narrow annulus model helps to increase the concentration. If the protoplanets at 
$\sim$1 au have roughly the lunar mass at $t_0$, however, they spread during accretion and the simulated Venus and Earth end up too
far from each other. The problems with radial mass spreading can be reduced if the initial protoplanets have roughly the Mars mass 
(Jacobson \& Morbidelli 2014) and started in a narrow annulus. 
\item Specifically, we find that 13\% of simulations in W11e/M20 (the narrow annulus model with 20 Mars-mass protoplanets; Jacobson
\& Morbidelli 2014) simultaneously satisfy the small Mars, $\Delta a$ and $\langle e,i \rangle$ 
constraints (Fig. \ref{mars1}). Adopting the Mars-mass protoplanets at $t_0$, however, limits the predictive power of the late-stage accretion 
model, because Mars is assumed to be fully formed at $t_0$. It also raises questions about whether even more massive planets could have formed 
at $\sim 1$ au before the gas disk dispersal. If the terrestrial planet system was practically in place at the time of the gas disk dispersal 
(e.g., formed by convergent migration of protoplanets in the gas nebula; Bro\v{z} et al. 2020), the significance of the late stage accretion 
would be relatively minor. 
\item We identified simulations where the masses and orbits of Venus, Earth and Mars are very well reproduced. In the highlighted case
from W11e/20M/{\tt case1}, two protoplanets grow near $r=1$ au to $M \sim 0.5$ $M_{\rm Earth}$ and collide at $t=41.3$ Myr after $t_0$. 
The timing and characteristics of this collision are consistent with the Moon-forming impact (Canup 2012, Kleine \& Walker 2017). 
With an eye to the Late Heavy Bombardment (LHB) of the Moon (Bottke \& Norman 2017), we will follow up on this work to extract 
model predictions for the history of planetesimal impacts on the terrestrial worlds. 

\end{enumerate}

\acknowledgements

\subsection*{Acknowledgements}
DN's work was supported by the NASA SSW program. FR's work was supported by the Brazilian National Council of Research (CNPq). 
R.D. acknowledges financial support from the NASA EW program. Resources supporting this work were provided by the NASA High-End Computing 
(HEC) Program through the NASA Advanced Supercomputing (NAS) Division at Ames Research Center. We thank the reviewer, Matthew Clement,
for many excellent suggestions to the submitted manuscript.

\clearpage

\begin{table}[t]
\begin{center}
\footnotesize
\begin{tabular}{lrrrrrrrrrrr}
\hline 
    & $f_{\rm g}$ & $S_{\rm c}$  &  $S_{\rm d}$   &  $\Delta a$  & $\langle e,i \rangle$ & $N_{\rm Mars}$ & $M_{\rm Mars}$ & $N_{\rm Merc}$ & $M_{\rm Merc}$ & 
$N_{\rm spl}$ & $f_{\rm ast}$ \\
                     &        &        &          &    (au)  &        &               & ($M_{\rm E}$) &      & ($M_{\rm E}$) & &  \\
solar system         &   --   & 89.9   & 0.0018   &  0.277  & 0.0274  &     1/1  & 0.107  &     1/1  & 0.055 &   0  &  --  \\
\hline
C98/{\tt nojov}      &   0.0  & 23.4   & 0.0022   &  --     & --      & 1.4/0.0  & 0.347  & 1.0/0.5  & 0.219 & 4.5  & 0.066 \\
C98/{\tt jovend}     &   0.4  & 34.1   & 0.0069   &  0.419  & 0.0866  & 0.8/0.5  & 0.262  & 0.5/0.3  & 0.226 & 1.5  & 0.019 \\
C98/{\tt case1}      &   0.5  & 30.0   & 0.0065   &  0.439  & 0.0584  & 1.0/0.4  & 0.150  & 0.8/0.4  & 0.203 & 0.4  & 0.008 \\
C98/{\tt case2}      &   0.1  & 22.1   & 0.0018   &  0.391  & 0.0810  & 1.0/0.0  & 0.379  & 1.0/0.0  & 0.270 & 1.0  & 0.007 \\
C98/{\tt case3}      &   0.3  & 28.7   & 0.0078   &  0.347  & 0.0938  & 2.3/0.3  & 0.085  & 1.0/0.3  & 0.247 & 1.0  & 0.012 \\
\hline
W11/{\tt nojov}      &   0.5  & 41.4   & 0.0002   &  0.368  & 0.0107 & 1.6/0.4  & 0.227  & 1.2/0.8  & 0.191 & 1.4 & 0.797 \\
W11/{\tt jovini}     &   0.5  & 39.6   & 0.0004   &  0.303  & 0.0181 & 1.4/0.2  & 0.520  & 0.8/0.4  & 0.201 & 0.6 & 0.782 \\
W11/{\tt jovend}     &   0.6  & 43.5   & 0.0015   &  0.348  & 0.0410  & 1.3/0.5  & 0.208  & 1.0/0.3  & 0.266 & 0.0  & 0.615  \\
W11/{\tt case1}      &   0.8  & 47.8   & 0.0029   &  0.342  & 0.0484  & 0.9/0.0  & 0.375  & 0.6/0.3  & 0.297 & 0.0  & 0.301  \\
W11/{\tt case2}      &   0.7  & 41.4   & 0.0056   &  0.379  & 0.0695  & 1.3/0.3  & 0.345  & 0.7/0.4  & 0.199 & 0.0  & 0.332  \\
W11/{\tt case3}      &   0.8  & 60.9   & 0.0024   &  0.375  & 0.0574  & 0.4/0.1  & 0.340  & 0.3/0.0  & 0.469 & 0.0  & 0.062  \\
\hline
L15/{\tt nojov}      &   0.4  & 24.4   & 0.0004   &  0.344  & 0.0215 & 1.3/0.0  & 0.451   & 1.8/0.8  & 0.221 & 2.5 & 0.343  \\
L15/{\tt jovini}     &   0.6  & 24.3   & 0.0008   &  0.289 & 0.0240 & 1.7/0.2   & 0.336   & 1.5/0.5  & 0.244 & 1.5 & 0.260  \\
L15/{\tt jovend}     &   0.7  & 26.1   & 0.0057   &  0.383  & 0.0558  & 1.1/0.3  & 0.365  & 0.9/0.0  & 0.349 & 0.4  & 0.489  \\
L15/{\tt case1}      &   0.1  & 28.6   & 0.0019   &  0.525  & 0.0216  & 1.0/1.0  & 0.151  & 1.0/1.0  & 0.150 & 0.0  & 0.270  \\
L15/{\tt case2}      &   0.5  & 24.0   & 0.0031   &  0.385  & 0.0476  & 1.0/0.0  & 0.373  & 1.2/0.2  & 0.389 & 0.6  & 0.287  \\
L15/{\tt case3}      &   0.3  & 36.9   & 0.0086   &  0.311  & 0.0916  & 0.3/0.3  & 0.066  & 1.3/0.3  & 0.403 & 0.0  & 0.050  \\
\hline
\multicolumn{12}{c}{{\it 100 simulation sets for} {\tt case1}} \\
C98s                 &   0.41 & 24.4   & 0.0069   &  0.375  & 0.0729  & 1.0/0.29 & 0.314 & 1.2/0.07 & 0.502 & 1.0  & 0.008 \\
C98s/25M             &   0.35 & 23.4   & 0.0053   &  0.382  & 0.0674  & 1.0/0.23 & 0.379 & 1.1/0.00 & 0.592 & 0.4  & 0.007 \\
W11e                 &   0.70 & 46.4   & 0.0040   &  0.354  & 0.0526  & 1.1/0.21 & 0.260 & 0.7/0.20 & 0.309 & 0.0  & 0.109 \\
W11e/20M             &   0.57 & 51.2   & 0.0034   &  0.349  & 0.0388  & 1.1/0.34 & 0.296 & 0.5/0.03 & 0.320 & 0.0  & 0.131 \\
\hline
\end{tabular}
\end{center}
\caption{Summary of the simulations performed in this work. The columns are: the (1) model name, (2) fraction of simulations
with good Venus/Earth counterparts ($f_{\rm g}$), (3) radial mass concentration ($S_{\rm c}$, Eq. (1)), (4) angular momentum 
deficit ($S_{\rm d}$, Eq. (2)), (5) radial separation between Venus and Earth ($\Delta a$), (6) mean orbital eccentricities and 
inclinations of Venus/Earth counterparts ($\langle e,i \rangle$), (7) number of Mars/strict Mars analogs per simulation 
($N_{\rm Mars}$), (8) average mass of Mars planets ($M_{\rm Mars}$), (9) number of Mercury/strict Mercury analogs per simulation 
($N_{\rm Merc}$), (10) average mass of Mercury planets ($M_{\rm Merc}$), (11) number of planets ($M>0.01$ $M_{\rm Earth}$) 
stranded in the asteroid belt at $t=200$ Myr ($N_{\rm spl}$), (12) surviving fraction of main-belt asteroids 
($f_{\rm ast}$). Columns (3)-(6), (8) and (10)-(12) give numbers averaged over all simulations that produced good Venus/Earth counterparts
($f_{\rm g}$ in column (2) and the main text). For the C98/{\tt nojov} model, where $f_{\rm g}=0$, the values are averaged 
over all ten simulations. The C98/{\tt jovini} model --not shown here-- did not produce useful results.}
\end{table}

\clearpage

\clearpage

\clearpage

\clearpage

\clearpage

\clearpage

\clearpage

\clearpage

\clearpage

\clearpage

\clearpage


\clearpage

\clearpage

\clearpage

\clearpage

\clearpage

\clearpage

\clearpage

\clearpage

\clearpage

\clearpage

\end{document}